\documentclass[apj]{emulateapj}

\newcommand{\msun}{$M_{\odot}$}

\newcommand{\zsun}{$Z_{\odot}$}
\newcommand{\rxjfull}{RX~J0152.7-1357}
\newcommand{\msfull}{MS~1054.4-0321}  
\newcommand{\rxj}{RX~J0152-13} 

\newcommand{\rtwoN}{$R_{200,N}^p$}
\newcommand{\rtwoS}{$R_{200,S}^p$}

\newcommand{\rtwoNval}{0.76~Mpc}
\newcommand{\rtwoSval}{0.40~Mpc}

\newcommand{\galmpc}{Mpc$^{-2}$}
\newcommand{\pmpc}{Mpc$^{-1}$}
\newcommand{\kmps}{km~s$^{-1}$}

\newcommand{\ugz}{$u_z^{\prime}-g_z^{\prime}$}
\newcommand{\ug}{$u^{\prime}-g^{\prime}$}

\newcommand{\BVz}{$B_z-V_z$}
\newcommand{\Rz}{$R-z^{\prime}$}

\newcommand{\Vfil}{$V$}
\newcommand{\Rfil}{$R$}
\newcommand{\ifil}{$i^{\prime}$}
\newcommand{\zfil}{$z^{\prime}$}
\newcommand{\ufil}{$u^{\prime}$}
\newcommand{\gfil}{$g^{\prime}$}

\newcommand{\zspec}{$z_{\rm spec}$}
\newcommand{\zldp}{$z_{\rm LDP}$}
\newcommand{\OII}{[OII]~3727~\AA}
\newcommand{\zwindow}{$0.80<z<0.87$}
\newcommand{\mcut}{10.6}
\newcommand{\lomid}{11.0}
\newcommand{\midhi}{11.3}
\newcommand{\ntot}{754}  
\newcommand{\nmcut}{300}  
\newcommand{\nldp}{649}   
\newcommand{\nldponly}{475}
\newcommand{\nldpspecz}{174}
\newcommand{\nspeczonly}{105}
\newcommand{\nspec}{279}     
\newcommand{\mcutrange}{$M>4 \times 10^{10}$~\msun\ ($\log M/M_{\odot} >~$\mcut)}
\newcommand{\lorange}{\mcut$~< \log M/M_{\odot} <~$\lomid}
\newcommand{\midrange}{\lomid$~< \log M/M_{\odot} <~$\midhi}
\newcommand{\hirange}{$\log M/M_{\odot} >~$\midhi}

\begin{document}

\title{A Wide-Field Study of the $\lowercase{z} \sim 0.8$ Cluster \rxjfull: the Role of Environment in the Formation of the Red-Sequence \altaffilmark{1,2,3}}

\author{Shannon G. Patel\altaffilmark{4},
Daniel D. Kelson\altaffilmark{5},
Bradford P. Holden\altaffilmark{4},
Garth D. Illingworth\altaffilmark{4},
Marijn Franx\altaffilmark{6},
Arjen van der Wel\altaffilmark{7},
Holland Ford\altaffilmark{7}
} 

\altaffiltext{1}{This paper includes data gathered with the 6.5~meter Magellan Telescopes located at Las Campanas Observatory, Chile.}
\altaffiltext{2}{Based in part on data collected at Subaru Telescope, which is operated by the National Astronomical Observatory of Japan.}
\altaffiltext{3}{Some of the data presented herein were obtained at the W. M. Keck Observatory, which is operated as a scientific partnership among the California Institute of Technology, the University of California and the National Aeronautics and Space Administration. The Observatory was made possible by the generous financial support of the W.M. Keck Foundation.}
\altaffiltext{4}{UCO/Lick Observatory, University of California, Santa Cruz, CA 95064; patel@ucolick.org}
\altaffiltext{5}{Observatories of the Carnegie Institution of Washington, Pasadena, CA 91101}
\altaffiltext{6}{Leiden Observatory, Leiden University, P.O. Box 9513, NL-2300 AA Leiden, Netherlands}
\altaffiltext{7}{Department of Physics and Astronomy, Johns Hopkins University, 3400 North Charles Street, Baltimore, MD 21218}

\begin{abstract}
We present the first results from the largest spectroscopic survey to date of an intermediate redshift galaxy cluster, the $z=0.834$ cluster \rxjfull.  We use the colors of galaxies, assembled from a $D \sim 12$~Mpc region centered on the cluster, to investigate the properties of the red-sequence as a function of density and clustercentric radius.  Our wide-field multi-slit survey with a low-dispersion prism (LDP) in the IMACS spectrograph at the 6.5~m Baade telescope allowed us to identify \nldponly\ new members of the cluster and its surrounding large-scale structure with a redshift accuracy of $\sigma_z/(1+z) \approx 1\%$ and a contamination rate of $\sim 2\%$ for galaxies with \ifil~$<23.75$~mag.  We combine these new members with the \nspec\ previously known spectroscopic members to give a total of \ntot\ galaxies from which we obtain a mass-limited sample of \nmcut\ galaxies with stellar masses \mcutrange.  We find that the red galaxy fraction is $93 \pm 3\%$ in the two merging cores of the cluster and declines to a level of $64 \pm 3\%$ at projected clustercentric radii $R \ga 3$~Mpc.  At these large projected distances, the correlation between clustercentric radius and local density is nonexistent.  This allows an assessment of the influence of the local environment on galaxy evolution, as opposed to mechanisms that operate on cluster scales (e.g. harassment, ram-pressure stripping).  Even beyond $R>3$~Mpc we find an increasing fraction of red galaxies with increasing local density.  The red galaxy fraction at the highest local densities in two large groups at $R>3$~Mpc matches the red galaxy fraction found in the two cores.  Strikingly, galaxies at intermediate densities at $R>3$~Mpc, that are not obvious members of groups, also show signs of an enhanced red galaxy fraction.  Our results point to such intermediate density regions and the groups in the outskirts of the cluster, as sites where the local environment influences the transition of galaxies onto the red-sequence.
\end{abstract}


\section{Introduction}
The environment of a galaxy has long been known to influence its properties, such as morphology, color, and star formation.  Numerous investigations showed that galaxy populations in rich cluster environments were dominated by red elliptical and S0 galaxies \citep{wolf1901, smith1935, zwicky1942, oemler1974}.  \citet{oemler1974}, in particular, postulated that more evolved clusters hosted a larger proportion of early-type galaxies, suggesting a link between galaxy properties and the buildup of clusters.  \citet{dressler1980} subsequently studied a diverse set of clusters and made the key discovery that at a fixed local density, relaxed and un-relaxed clusters host a similar fraction of early-type galaxies and that this fraction increased at higher local densities.  Dressler's morphology-density relation (MDR) challenged the long-standing role of ram-pressure stripping \citep{gunn1972} in transforming field spirals into cluster early-types galaxies, and instead pointed to the importance of local processes.  In recent years, studies have probed into the lower density environments that are characteristic of groups \citep{postman1984,zabludoff1998} and the field galaxy population  \citep{kauffmann2004, baldry2006}, and found the local environment to be an important factor there as well.  In general, these studies found the {\em lowest} density environments to be comprised of lower mass, blue, late-type galaxies, a stark contrast to the types of galaxies found at higher densities.

Studies of galaxy properties and their dependencies on local environment are key to understanding the mechanisms that are responsible for galaxy evolution.  For example, in the high density cluster environment processes like ram pressure stripping \citep{gunn1972} and harassment \citep{moore1996} may suppress star formation and consequently evolve galaxies towards the red-sequence.  At intermediate densities that are characteristic of the group environment, galaxy-galaxy interactions and mergers are more probable, given the favorable relative velocities of galaxies \citep{cavaliere1992}.   Early HST observations of distant clusters revealed tidal interactions of blue disk galaxies, implying a strong role for galaxy-galaxy interactions in forming present day cluster ellipticals \citep{dressler1994,dressler1994b}.  An analysis of galaxy properties in different environments can therefore shed light on the processes responsible for the observed buildup of the red-sequence between $z \sim 1$ and today \citep{bell2004b, faber2007}.  For example, given that cluster galaxies comprise a small fraction of their $z \sim 1$ sample, \citet{cooper2006,cooper2007} argue against processes such as ram pressure stripping and harassment as the sole origin of the observed color-density trend found at low redshift.  Instead, they suggest that processes that form the high early-type fraction in $z \sim 0$ groups of galaxies dominate red-sequence formation at higher redshifts.

The outskirts of {\em massive} clusters present an ideal location to study the impact of environment on galaxy properties over a broad range of local densities: from the high density cluster core to the intermediate density groups in the outskirts and the low density surrounding field.  Furthermore, the findings of \citet{holden2007} and \citet{vanderwel2007b} suggest that the morphological makeup of galaxies changes dramatically at the intermediate densities found in the outskirts of clusters.  These mass-limited studies found that while in the cores of clusters the fraction of early-type galaxies is high and does not evolve significantly between $z \sim 1$ and $z=0$, the early-type fraction (ETF) for galaxies in the low density field is lower and also does not evolve.  Because clusters grow by accreting galaxies from the field, and massive clusters are expected to grow by a factor of $\sim 2-3$ in mass between $z \sim 1$ and $z=0$ \citep{wechsler2002}, the large reservoir of infalling galaxies from the field must transform in the outskirts of clusters in order to preserve the high ETF found in the high density core.  Thus, the outskirts of massive clusters provide an opportunity to assemble a large sample of galaxies in environments where we expect transformations to be occurring.

Studying the properties of galaxies in the environments of clusters at intermediate redshifts has several advantages.  At such redshifts, around $z \sim 0.8$, at a lookback time of $\sim 7$~Gyr, a number of clusters exist where the formation of a massive cluster out of the large scale structure can be witnessed.  This is an epoch of rapid buildup in the cluster population.  Observations of the evolution of the cluster mass function indicate that the {\em number} of massive clusters increases by a factor of a few since such redshifts \citep{vikhlinin2008}.  In addition, as noted above, simulations suggest that individual clusters grow in {\em mass} by a factor of $\sim 2-3$ over the same time period \citep{wechsler2002}.  This changing landscape therefore allows one to study the properties of galaxy populations that will transform, evolve, and assemble into the present day cluster population.  On much smaller physical scales, the star forming activity in galaxies was also different at these intermediate redshifts.  In general, galaxies formed stars at a higher rate in the past \citep{lilly1996, madau1998}, particularly higher mass galaxies that will have ceased star formation by the present epoch \citep{cowie1996, noeske2007b}.  {\em At a redshift of $z \sim 0.8$, the contrast between star-forming and non-star-forming galaxies is accentuated, allowing us to better highlight those environments that influence star formation.}

We focus in this paper on a mass-limited sample for this important epoch of cluster mass growth.  Luminosity-limited samples selected in the rest-frame UV include many bright, low mass, star forming galaxies as a result of the large scatter in their mass-to-light ratios at a fixed galaxy stellar mass.  Selecting by stellar mass thus provides a more well-defined and homogeneous sample.

In this paper, we study the colors of galaxies in the $z=0.834$ cluster \rxjfull\ (hereafter \rxj) and its outskirts.  This is a very rich cluster with extensive imaging and spectroscopic datasets that now cover a very wide area (out to $\sim 6$~Mpc in radius).  As such it is an ideal target for a study of galaxy evolution and transformation over a wide range of densities, from the core to the field.  Our initial spectroscopic sample of members is currently $\sim 750$ objects.  These galaxies are the focus of our first paper on this cluster.  Such samples are becoming large enough to provide robust constraints on cluster galaxy properties (e.g. mass functions, star formation rates, morphologies, environmental trends, etc).  The use of spectroscopic redshifts, as opposed to photometric ones, ensures that our membership list, and therefore results, do not suffer from the systematics that trouble photometric surveys.  Ultimately we expect to reach 2000-3000 spectroscopic members between two clusters at $z \sim 0.8$.  The remarkable feat employs the new low-dispersion prism (LDP) in the Inamori Magellan Areal Camera and Spectrograph (IMACS) on the Baade 6.5~m telescope at Magellan (see below for more details), further enhancing our insight into galaxy evolution in a variety of environments at $z \sim 0.8$.

In \S~\ref{observations} we describe the data employed in this wide-field observing campaign.  In \S~\ref{analysis} the derived properties of galaxies in our sample are discussed.  Galaxy colors and their trends with clustercentric radius and local density are presented in \S~\ref{results}.  The implications of these results are discussed in \S~\ref{discussion}.  We summarize our findings in \S~\ref{summary}.

Throughout this paper, magnitudes from Sloan filters will be on the AB system, and Johnson filters on the Vega system.  The use of both systems is a consequence of adhering to convention and making comparisons to previous work more manageable.  We assume a $\Lambda$CDM cosmology with $H_0=70$~\kmps\ \pmpc, $\Omega_{M}=0.30$, and $\Omega_{\Lambda}=0.70$, resulting in a look-back time of 7.0~Gyr and an angular scale of 0.46~Mpc~arcmin$^{-1}$ at the cluster redshift, $z=0.834$.

\section{Data} \label{observations}

\begin{figure*}
\plotone{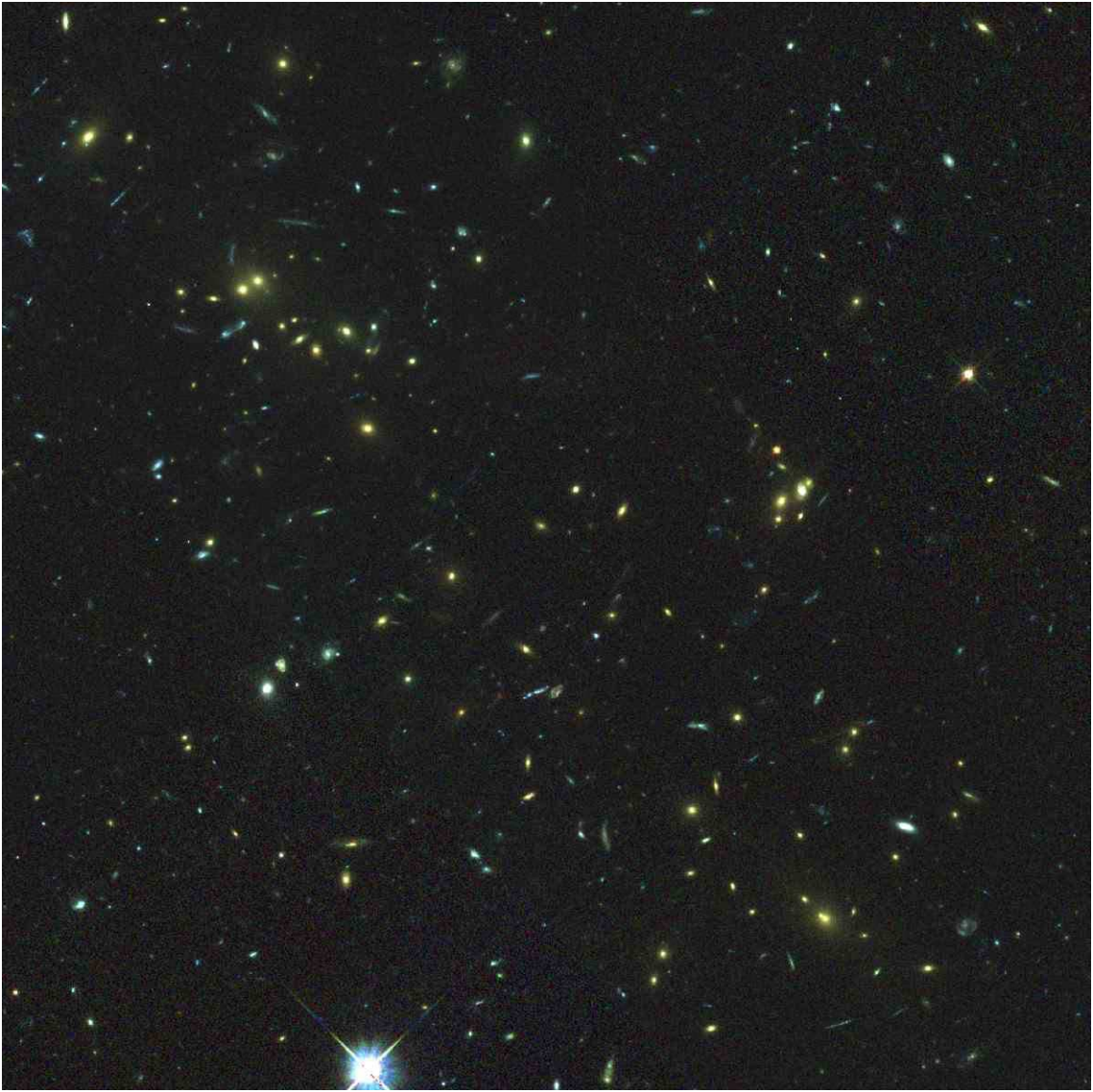}
\caption{Color image of a 1.9\arcmin$\times$1.9\arcmin\ (0.9~Mpc~$\times$~0.9~Mpc) region surrounding the two cores of \rxj\ constructed with ACS F625W ($r^{\prime}$), F775W (\ifil), and F850LP (\zfil) filters.  The more massive northern core is located near the upper left corner of the figure while the southern core is located near the bottom right.  The 3-D virial radii ($r_{200}$) for the northern and southern cores, as computed from their velocity dispersions, are 1.2~Mpc and 0.6~Mpc respectively (see \S~\ref{virialradii}).  The presence of two distinct cores highlights the fact that \rxj\ is still in the process of forming at this epoch.} \label{rxjcolorim}
\end{figure*}

\begin{figure*}
\plotone{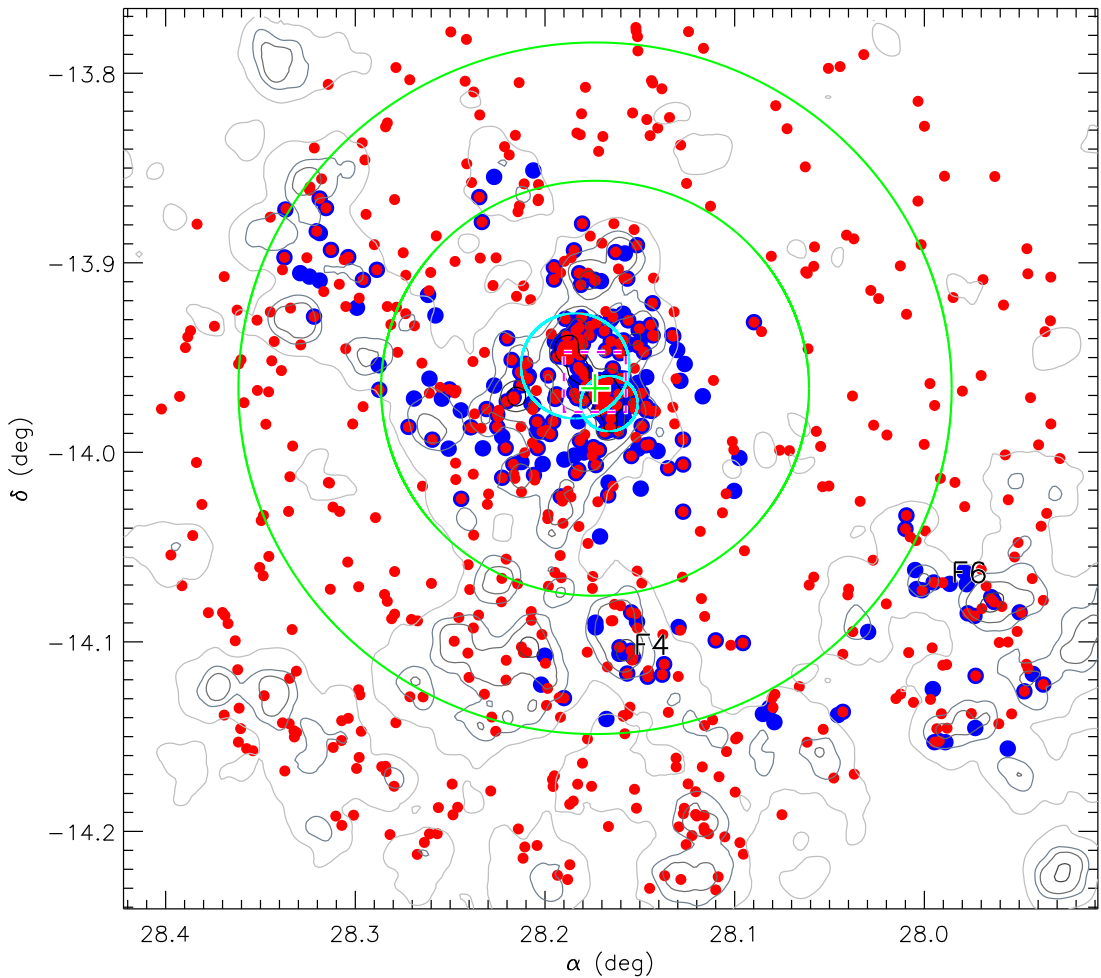}
\caption{Spatial distribution of galaxies in and around the $z=0.834$ cluster \rxj.  Blue circles are previously known spectroscopic members from high-resolution spectroscopy, while red circles are from IMACS LDP spectroscopy.  All the redshifts are in the range $0.80<z<0.87$ and have \ifil$~<23.75$~mag.  An HST ACS color image of the region within the dashed magenta colored square is shown in Figure~\ref{rxjcolorim}.  The two cyan colored circles in the center represent the projected virial radii for the northern and southern cores, \rtwoN~=~\rtwoNval\ and \rtwoS~=~\rtwoSval, respectively.  The superscripted `$p$' indicates that these virial radii account for a spherical projection factor of $2/\pi$.  The center of the overall cluster system is indicated with a plus ($+$) and the two outer green circles represent clustercentric radii of 3 and 5~Mpc.  Local density contours are shown in gray for $\Sigma =$ 10, 20, 30, 60, and 100~\galmpc.  Two groups in the outskirts identified by \citet{tanaka2006}, the F4 and F6 group, are labeled.  Our wide-field survey of \rxj\ has revealed a diversity of environments in the outskirts of the cluster.} \label{map0152}
\end{figure*}

\rxj\ was discovered as part of the WARPS survey with {\em ROSAT} \citep{scharf1997} and was also independently identified in the RDCS \citep{rosati1998}.  Kinematic \citep{demarco2005,girardi2005} and X-ray \citep{maughan2003,maughan2006} studies suggest significant substructure in \rxj, most notably the presence of two distinct cores separated spatially by 1.5\arcmin\ and in redshift by $\Delta z = 0.01$ ($\Delta v \sim 1600$~\kmps).  A color image of the cluster created with HST ACS imaging using the F625W, F775W, and F850LP filters is shown in Figure~\ref{rxjcolorim} \citep[see][for a description of these data]{blakeslee2006}.  Mass estimates for the combined system differ based on the assumptions made and typically range from $(3.6 \pm 1.0) \times 10^{14}$~\msun\ \citep{demarco2005} to $(1.1 \pm 0.2) \times 10^{15}$~\msun\ \citep{maughan2003}.  \citet{tanaka2005, tanaka2006} identified several groups in the outskirts of the cluster, some of which will fall in by $z=0$.  Those galaxies in such high density regions in the cluster outskirts will be discussed in later sections.  \citet{blakeslee2006} measured the mean redshift of all early-type galaxies in the two cores of \rxj\ to be $z=0.834$.  We assume this value to represent the redshift of the cluster.

\subsection{Imaging} \label{imaging}
We used archival Subaru Suprime-Cam imaging \citep{miyazaki2002b} of \rxj\ \citep{kodama2005} in \Vfil \Rfil \ifil \zfil\ for photometry and spectroscopic target selection.  The Suprime-Cam imaging was made up of 10 pointings covering a $\sim$29\arcmin~$\times$~39\arcmin\ field of view.  The data were reduced using standard techniques for image reduction, including steps for overscan subtraction, flat-fielding, and stacking with cosmic-ray rejection.  The photometry was calibrated using frames obtained from the archive of standard stars taken on the same night.  \citet{kodama2005} measured a 5$\sigma$ limiting \ifil-band depth for a $D=2$\arcsec\ aperture to be 26.1~mag, which is significantly deeper than our spectroscopic selection (\S~\ref{spectroscopy}).  We used SExtractor \citep{bertin1996} to identify objects and DAOPHOT II \citep{stetson1987} to perform circular aperture photometry because of its superior estimate for the sky background.  Colors were defined with a fixed circular aperture size of $D=1$\farcs2, which was twice the seeing FWHM.  We also used a fixed aperture size of $D=3$\arcsec\ for total magnitudes.  Given the 0.6\arcsec\ seeing, the $D=3$\arcsec\ aperture encompassed most of the light ($\sim 90\%$) for the typical galaxy at the cluster redshift in the Suprime-Cam imaging.  For the brightest galaxies this may slightly underestimate luminosities and therefore stellar masses, but not enough to exclude them from our mass-limited sample (\S~\ref{masses}).

\subsection{Spectroscopy} \label{spectroscopy}
The sample of galaxies discussed in this paper incorporates cluster members with spectroscopic redshifts from the catalogs of \citet{demarco2005} and \citet{tanaka2006}.  These observations focused on the core of the cluster and groups in the outskirts.  Additional members with redshifts from Keck and Magellan spectroscopy were also included.  Galaxies with redshifts from these high-resolution spectroscopy catalogs in the range \zwindow\ are shown as blue circles in Figure~\ref{map0152}.

To expand the spectroscopic sample of galaxies in the vicinity of \rxj, we employed a unique instrument that allowed us to gather $\sim 2000$ redshifts per pointing over a $\sim$27~arcmin diameter field of view ($\sim$12~Mpc at the cluster redshift): a low-dispersion prism (LDP; built for the PRIMUS collaboration by S. Burles, private communication) in the Inamori Magellan Areal Camera and Spectrograph \citep[IMACS;][]{bigelow2003, dressler2006} on the Baade 6.5~m telescope at Magellan.  In this paper we present results from the first set of observations, which have already allowed us to build one of the largest {\em spectroscopic} samples of galaxies in a cluster and its outskirts (now \ntot\ members) at intermediate redshift.  Over subsequent seasons, we expect to obtain 2000-3000 members in and around two clusters at $z \sim 0.8$.  

With the LDP, IMACS provided a spectral resolving power of $80 > R > 10$ over a corresponding wavelength range of 4500~\AA $~< \lambda <~$ 9000~\AA.  The data were taken using nod \& shuffle for sky subtraction \citep{glazebrook2001} with the slits being just long enough to accommodate two object spectra (A and B) with each nod.  Wavelength calibration was performed using Helium lamp exposures taken between science frames, providing nearly a dozen features from 3888~\AA\ to 10,830~\AA.  The positions of the lines were solved for simultaneously in order to account for blending.  We then constructed mappings of wavelength to position on the CCD using trivariate polynomials of slit mask position and wavelength.  One-dimensional spectra were constructed by simultaneously extracting the A and B pairs using an optimal extraction technique similar to \citet{horne1986}.  The spectra were flux calibrated.  We used a total of 5 slitmasks, each with $\sim$2000 slits.  The first 3 masks selected galaxies from our Suprime-Cam photometry catalog with \ifil~$<$~23~mag, while the last 2 went deeper to \ifil~$<$~23.75~mag.  Roughly $\sim 75\%$ of the objects had $>2$~hrs of exposure and $\sim 50\%$ had $>3$~hrs of exposure.  The cumulative time of observing all 5 masks would have been equivalent to about 3 nights on the telescope.  The efficiency of this setup is clear and allows us to obtain an order of magnitude more redshifts per night compared to traditional multi-object spectroscopy.  We stress that {\em no priors, such as red-sequence membership or photometric redshifts, were implemented in our spectroscopic selection, thereby minimizing biases in our sample}.  Observations with the LDP are ongoing and we therefore plan to discuss a larger and more complete sample of cluster galaxies in a future paper.

\section{Analysis} \label{analysis}

\subsection{Prism Redshifts} \label{redshifts}

\begin{figure}
\epsscale{1.2} 
\plotone{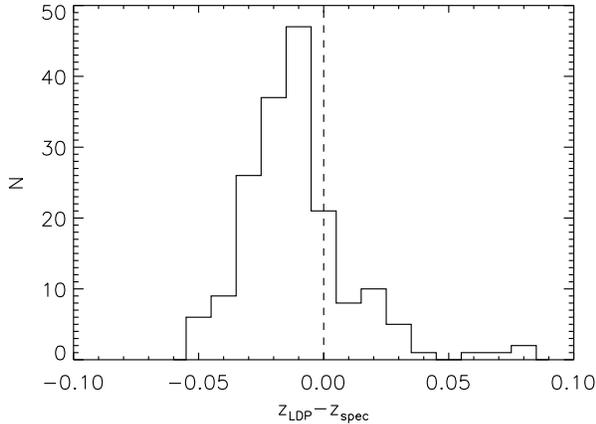}
\caption{For members of the cluster and its outskirts, a histogram of the deviation of LDP redshifts (\zldp) from redshifts measured from higher resolution spectra (\zspec).  The biweight computed standard deviation of $z_{\rm LDP}-z_{\rm spec}$ is $\sigma = 0.018$ and the offset from $z_{\rm LDP}-z_{\rm spec} = 0$ is quite small, with $\Delta z = -0.013$.  The good agreement between \zspec\ and \zldp\ establishes the LDP in IMACS as a powerful tool in studying large quantities of galaxies at intermediate redshift.} \label{redshifts_hist}
\end{figure}

Until now, surveys of clusters at high redshift have relied, to some degree, on photometric redshifts (photo-z's) to build samples large enough to conduct studies with adequate statistics \citep{tanaka2005,delucia2007b}.  Such studies have suggested that selection effects associated with photo-z's must be considered when interpreting results.  For instance, \citet{tanaka2006} show that photo-z's for blue galaxies are systematically lower than their spectroscopic values, and in the case of \rxj, results in a deficit of such galaxies in the cluster.  With the LDP, we are able to avoid both the systematics and the catastrophic redshift failures that complicate photometric studies.  A detailed discussion of the methodology and errors of LDP redshift determinations will be presented in a future paper (Patel et al. 2009, in preparation) but is summarized here.

The LDP spectra are low resolution grism spectra taken through short slitlets. The advantage of this observing technique is that spectra for very large samples of galaxies can be obtained quickly.  The resolution varies from $\sim 50$~\AA\ in the blue to $\sim 900$~\AA\ in the red (for FWHM~$\sim 4$~pixels).  Such low resolution requires non-standard data reduction and processing in order to recover accurate redshifts.  However, as shown below through comparison with a substantial sample of galaxies observed spectroscopically at higher resolution, this technique has proven to return large numbers of spectra quickly and reliably, and has great potential for future studies of intermediate and high redshift galaxies.

Redshifts for the LDP sample of galaxies were determined by simultaneously fitting the Suprime-Cam photometry and LDP spectrophotometry with a grid of template spectra generated using \citet[BC03]{bc03}.  The templates used constant star formation models with the onset of star formation occurring at $z=5$.  The grid of template spectra was constructed as a function of 3 parameters: redshift, metallicity, and truncation time.  The truncation times span a wide range in order to mimic galaxies that are passively evolving to constantly star-forming.  The fitting included a non-negative Gaussian component at the position of \OII.  At each grid point, a non-negative least squares algorithm was used to fit the stellar and \OII\ components to the observed spectral energy distribution (SED) of a given galaxy.  The quality of the fit at each grid point was determined by $\chi^2$.  The template spectra were convolved with the prism's wavelength-dependent resolution and dispersion prior to fitting the observations.  We note that more recent trials of SED fitting have tested the inclusion of varying amounts of extinction due to dust but no discernible difference in redshifts have been found.  We plan to discuss \OII\ measurements from the SED fitting in a future paper.  

We use a likelihood analysis to identify members of the \rxj\ superstructure. The redshift likelihood function for each galaxy is defined such that the probability density at a given redshift, $p(z)$, is proportional to $\chi^2$ of the template fit to the SED in the following way: $p(z) \propto \exp(-\chi^2/2)$, and the likelihood for a galaxy to be in the redshift range $z+dz$ is given by $\propto p(z)dz$.  The redshift likelihood function is then used to find all galaxies with $>$50\% of their likelihood within the redshift interval \zwindow.  This redshift interval has been used in previous studies to discuss galaxies in \rxj\ and its outskirts \citep{demarco2005, blakeslee2006}.  LDP galaxies with redshifts in the range \zwindow\ are shown as red circles in Figure~\ref{map0152}.

We found \nldp\ galaxies in \rxj\ and its outskirts with the LDP out of 7659 extracted spectra from the 5 masks (roughly $75\%$ of these 7659 had well constrained redshifts).  The median value for the full width of the 95\% redshift confidence interval for the cluster members is $\delta z_{95} = 0.060$, corresponding to a 1-$\sigma$ uncertainty of $\sigma = 0.015$.  Comparing \nldpspecz\ LDP redshifts, \zldp, to those derived from higher resolution spectra, \zspec, we find a scatter of $\sigma = 0.018$, implying that our error estimates for LDP redshifts are robust.  The histogram of $z_{\rm LDP}-z_{\rm spec}$ is shown in Figure~\ref{redshifts_hist}.  Note that there is a small, but systematic offset from $z_{\rm LDP}-z_{\rm spec} = 0$ of $\Delta z = -0.013$, which is accounted for in assigning membership.  The scatter for red-sequence galaxies, $\sigma = 0.015$, is slightly smaller than galaxies with bluer colors, $\sigma = 0.019$.  Of the 174 with measurements of both \zldp\ and \zspec, 4 (2.3\%) have \zspec\ outside of the cluster redshift interval but \zldp\ inside of it. However, none of these 4 galaxies deviate from the boundaries of the cluster interval by more than $\Delta z \approx 0.06$.  Thus, the catastrophic redshift failures known to trouble photometric redshift samples are substantially reduced for LDP-selected members.  In addition, we expect the agreement between \zldp\ and \zspec\ to improve in the future as we refine the reduction and fitting of LDP spectra and track down systematic sources of uncertainty.

We incorporate \nspeczonly\ additional members, from existing spectroscopic catalogs, that have not yet been observed with the LDP, to build a sample of \ntot\ galaxies at the redshift of the cluster with \ifil$~<~23.75$~mag.

\subsection{Spectroscopic Completeness} \label{completeness}

\begin{figure}
\epsscale{1.2} 
\plotone{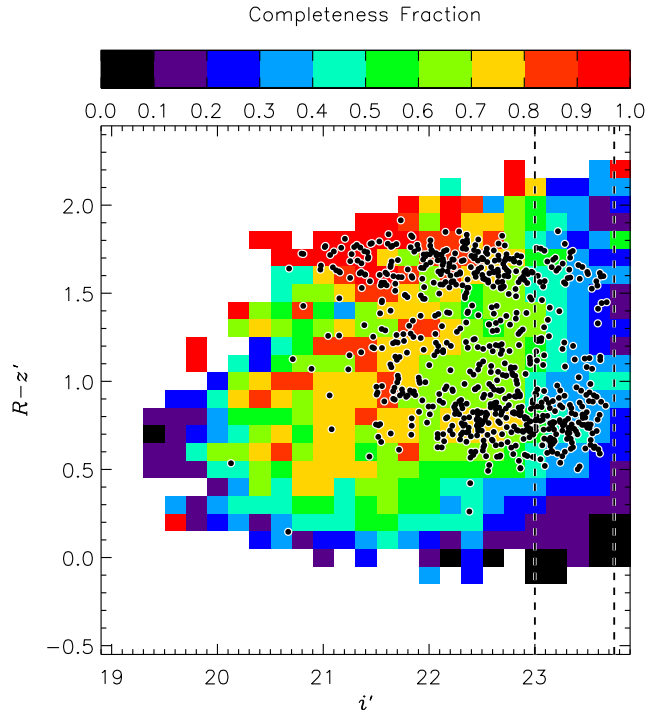}
\caption{Spectroscopic completeness of our survey of \rxj\ as a function of \Rz\ and \ifil\ (Note that the \Rz\ color represents a Vega magnitude minus an AB magnitude.  See \S~\ref{kcorrection} for a conversion between the two systems for the \Rfil\ filter).  The completeness fraction in a given color-magnitude bin is given by $N_z/N$, where $N_z$ is the number of galaxies with measured redshifts and $N$ is the total number of galaxies targeted from the Suprime-Cam photometry catalog with \ifil~$<23.75$~mag.  Note that the first 3 slitmasks selected galaxies with \ifil~$<23$~mag while the last 2 selected galaxies down to the fainter limit of \ifil~$<23.75$~mag.  The vertical dashed lines indicate these magnitude limits.  The \ntot\ galaxies defined to be members of the cluster superstructure are overplotted.  Note that the spectroscopic completeness for galaxies on the red-sequence near \Rz~$\sim 1.6$ is above $\sim 50\%$ for \ifil~$<23$~mag.  At brighter magnitudes along the red-sequence, the spectroscopic completeness quickly ramps up to $>80\%$.  Overall, our spectroscopic completeness is high and the variations in this map are accounted for in our calculations in order to minimize any biases in our results.} \label{fig_completeness}
\end{figure}

To correct for the non-uniform depth of the selection between the first and second sets of spectroscopic data, we constructed the spectroscopic completeness as a 2-D function of \ifil\ and \Rz.  This is shown in Figure~\ref{fig_completeness}.  Galaxies are binned in \ifil\ in increments of 0.2~mag and in \Rz\ by 0.1~mag.  The completeness fraction for a particular color and magnitude bin is given by $N_{z}/N$, where $N_z$ is the number of galaxies with sufficient constraints on their redshifts to assign membership and $N$ is the number of galaxies in the object catalog with \ifil$~<23.75$~mag.  This magnitude limit for the spectroscopic target selection is much brighter than the photometric limit of the Suprime-Cam catalog.  In our analysis below we weigh each galaxy with the inverse of the completeness fraction in its color-magnitude bin.

\subsection{Rest-frame Colors and Magnitudes} \label{kcorrection}

We transform colors and magnitudes to the rest-frame using the same method as \citet{blakeslee2006} and \citet{holden2006}.  Using a range of $\tau$-model spectra from BC03, we calculate magnitudes in the rest-frame for the relevant filters used in this paper (\ufil\gfil$BV$) and in the observed redshifted frame using Suprime-Cam \Vfil\Rfil\ifil\zfil.  While these models are different from the constant star forming models used to fit SEDs and obtain redshifts, their use makes our rest-frame colors and magnitudes consistent with previous work.  Also, our results are not significantly affected by this choice.  The LDP SEDs themselves are not used because they are lower S/N than the Suprime-Cam photometry and contain systematic uncertainties in their flux calibration.  In creating our $\tau$-models, we used metallicities of 0.4, 1.0, and 2.5 \zsun, $e$-folding timescales $\tau$ ranging from 0.1 to 5~Gyr, and elapsed time since onset of star formation ranging from 0.5 to 6~Gyr.  These models span the colors seen in our observations for cluster members.  Suprime-Cam \Vfil\Rfil\ifil\zfil\ filter transmissions, multiplied by the quantum efficiency of the detector, were used for the observed transmission curves and Buser's $B3$ and \Vfil\ \citep{buser1978} and SDSS \ufil\gfil\ transmission curves \citep{fukugita1996} supplied with BC03 were used for the rest-frame.  The best-fit transformations from observed to rest-frame magnitudes and colors are given below.  We follow the convention where rest-frame magnitudes and colors are subscripted with the letter `$z$'.
\begin{eqnarray}
B_z &=& i^{\prime}-0.349(i^{\prime}-z^{\prime})+0.797-(0.027) \nonumber\\
V_z &=& z^{\prime}-0.368(i^{\prime}-z^{\prime})+0.703-(0.018) \nonumber\\ 
B_z-V_z &=& 1.019(i^{\prime}-z^{\prime})+0.093-(0.009) \nonumber\\
u_z^{\prime}-g_z^{\prime} &=& 0.835(R-z^{\prime})+0.239-(0.014)
\end{eqnarray}

These transformations have a 1-$\sigma$ scatter of 0.007, 0.008, 0.012, and 0.016~mag respectively.  Note that in the conversion to \ugz, the \Rfil~$-$~\zfil\ color represents a Vega magnitude minus an AB magnitude.  To aid the reader, the conversion between the two systems for the Suprime-Cam \Rfil\ filter is $R_{\rm AB} = R_{\rm Vega} +0.185$.  The last term in each transformation, given in parentheses, corrects the derived rest-frame magnitudes for Galactic extinction using the dust maps of \citet{schlegel1998}.  We use $B_z$ and \BVz\ to compute stellar masses in \S~\ref{masses} and $V_z$ for selecting galaxies to compute local densities in \S~\ref{localdensities}.  The \ugz\ color is used to study stellar populations since this combination of filters straddles the 4000~\AA\ break and is therefore sensitive to recent or ongoing star formation.  Our rest-frame \ugz\ colors show a scatter of $\sim$0.04~mag when compared to those derived from \citet{holden2007} who use HST ACS data.  Although different apertures were used in the two works to compute colors, the quoted scatter should provide a reasonable {\em conservative} estimate of the error in colors, particularly since \citet{vandokkum1998} found that color gradients were very small at these redshifts.  We therefore adopt $\sim$0.04~mag as the uncertainty in \ugz\ colors.

All quantities in this paper have been calculated assuming the redshift for all members is $z=0.834$.  However, the actual redshifts range over \zwindow.  Therefore properties such as rest-frame color and stellar mass (\S~\ref{masses}) will deviate slightly from those values derived using the transformations given above \citep[also see][]{blakeslee2006}.  We find {\em maximum} deviations, resulting from the range in redshifts, in rest-frame \ugz\ color of $\sim 0.05$~mag and stellar mass of $\sim 0.05$~dex.

\subsection{Stellar Masses} \label{masses}

\begin{figure}
\epsscale{1.25} 
\plotone{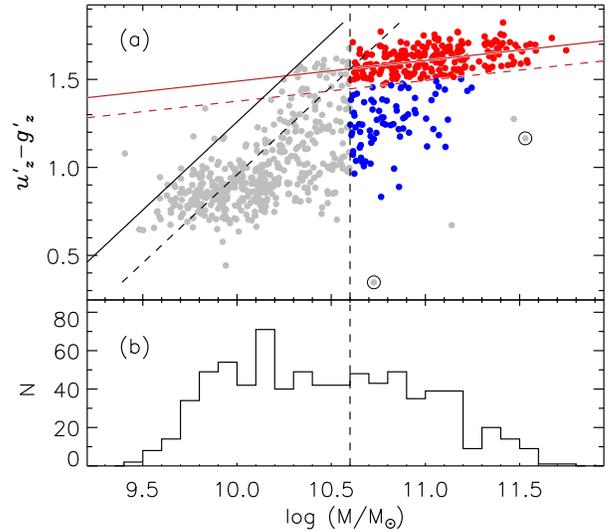}
\caption{Color vs. stellar mass for galaxies in \rxj\ and its outskirts (a).  The black diagonal dashed and solid lines represent the derived limits of our spectroscopic selection criteria of \ifil~=~23~mag and \ifil~=~23.75~mag respectively.  The dashed vertical line indicates our mass cut of \mcutrange\ above which we are $>$50\% complete.  This completeness limit can be seen on the red-sequence in Figure~\ref{fig_completeness} at \ifil~$\sim 23$~mag.  The solid red line represents the color-mass relation (CMR) for red galaxies (see \S~\ref{colorfractions}).  The 2$\sigma$ scatter for this relation separates red galaxies from blue ({\em dashed red line}).  The 2 known AGN detected in X-ray observations are circled \citep{demarco2005}.  Gray data points above the mass limit are ignored in our analysis because their colors or magnitudes are contaminated by nearby stars.  The histogram of $\log M/$\msun\ (b) shows the distribution of stellar masses, derived using the \citet{bell2003} relation between rest-frame $B-V$ color and $B$-band mass-to-light ratio.  Our mass-limited sample consists of \nmcut\ galaxies, $73\%$ of which are red.  This dataset provides a large, statistically robust sample from which we can assess the role of environment in evolving massive galaxies.}\label{colormass}
\end{figure}

\citet{kelson2000c} found stellar mass-to-light ratios and rest-frame colors to be well correlated.  \citet{bell2001} explored the utility of this correlation for estimating the stellar masses of large samples of galaxies with a more formalized framework in \citet{bell2003}.  For this paper we therefore compute stellar masses for galaxies using the \citet{bell2003} relation between rest-frame $B-V$ color and $M/L_B$: $\log M/L_B=1.737(B-V)-0.942$.  \citet{bell2003} use a ``diet'' Salpeter IMF in computing $M/L$.  This IMF has the typical Salpeter slope ($\alpha=2.35$) between $0.35 < M/M_{\odot} < 125$ but is flat ($\alpha=0$) below $M< 0.35$~\msun.  This results in the same luminosity output at 70\% of the mass when compared to a traditional Salpeter IMF.  \citet{bell2003} provide conversions for $M/L$ derived from a diet Salpeter IMF to various other IMFs.  Assuming a distance modulus for \rxj\ of 43.61~mag, and $M_{B,\odot}=5.45$~mag, the mass estimates are given by the formula below.
\begin{eqnarray}
\log(M/M_{\odot}) &=& -0.4(M_{B_z}-M_{B,\odot}) + \nonumber\\
    && 1.737(B_z-V_z) - 0.942
\end{eqnarray}

According to \citet{bell2001}, $M/L$ uncertainties are on the order of 0.1-0.2~dex.  \citet{holden2006, holden2007} find a scatter of 0.25~dex about dynamical mass estimates for cluster galaxies at $z \sim 0.8$.  Because our masses were computed in a similar way, we assume this value represents our uncertainty in stellar masses.  The top panel of Figure \ref{colormass} shows the rest-frame \ugz\ color plotted against our derived masses.  The distribution of masses is given in the bottom panel.  For galaxies with mass \mcutrange, the sample is representative of the full object catalog with a uniform completeness of $>$50\%.  This value for our mass limit represents $\approx 0.5M^{\star}$ \citep[$M^{\star} \approx 10^{11}$~\msun,]{baldry2006}.  For the remainder of this paper, we focus on galaxies above this stellar mass limit, \mcutrange.

Finally, we would like to note that in this work quantities such as rest-frame color and mass-to-light ratio were not computed from the best-fitting truncation model template used to determine redshifts.  Instead, these quantities were computed from model star formation histories (SFHs) with exponentially declining star formation rates (i.e. tau-models).  This allowed comparison to previous work to be more accessible.  For this paper, we viewed the redshift determination to be an independent task and therefore chose a SFH for the model templates that: (1) provided accurate redshifts and (2) optimized computing time given the large amount of data that needed to be processed (truncated SFHs use one fewer free parameter than tau-models).  In future work, we may explore integrating the entire process of redshift determination and extraction of rest-frame quantities.

\subsection{Red Galaxy Fraction}  \label{colorfractions}
The colors of galaxies provide insight into the ages of their stellar populations.  Red galaxies are typically associated with old stellar populations, like those found in ellipticals and S0s.  Galaxies that actively form stars have blue colors and are typically comprised of spirals and irregular systems.  These late-type galaxies however may also appear red.  \citet{vanderwel2007b} find $38 \pm 8\%$ of field red-sequence galaxies at $z \sim 0.8$ with mass \mcutrange\ are late-types.  The presence of dust and the viewing angle often play a hand in the appearance of these reddened late-type galaxies, which might otherwise appear blue.  Galaxies with intermediate colors between the blue-red color bimodality have been linked to galaxies in transition from blue to red after star formation is quenched \citep{coil2008} but they are also known to be dust-obscured star forming galaxies \citep{bell2005, weiner2007}.

We define whether a galaxy is on the red-sequence based on its deviation from the color-mass relation (CMR) shown in Figure~\ref{colormass} (red line).  This CMR was derived by iteratively fitting a line to galaxies with mass \mcutrange\ and rejecting 2$\sigma$ outliers with each iteration.  The 2$\sigma$ scatter of non-rejected galaxies about the converged CMR was used to distinguish red and blue galaxies, with blue galaxies having \ugz\ colors $>2\sigma$ below the CMR (i.e. approximately $\sim 0.12$~mag or more below the ridge line).  This division between red and blue galaxies is shown in Figure~\ref{colormass} with a dashed red line.  ``Blue'' galaxies therefore include the blue cloud as well as any intermediate color objects.  Red galaxies account for $73\%$ of the mass-limited sample.

\subsection{Local Projected Galaxy Densities} \label{localdensities}
The local environment, or local galaxy density, has been shown to play a role in the evolution of galaxies \citep{baldry2006, cooper2007, holden2007}.  The effectiveness of different mechanisms responsible for such evolution often depends on the local environment.  To characterize the local environment for galaxies in our sample we computed the local projected galaxy density, $\Sigma$, at each galaxy location.

We use an $n$-th nearest neighbor scheme, where $n=7$, to compute $\Sigma$.  The local density at a given galaxy location is therefore given by
\begin{eqnarray}
\Sigma= \frac{n+1}{\pi d_7^2} \nonumber
\end{eqnarray}
where $d_7$ is the projected distance to the 7-th nearest neighbor.  Note that this is a physical density (as opposed to a comoving density).

Given the potential for large spatial variations in the completeness of our spectroscopic survey, we used a multi-color-selected list of galaxies compiled from the Suprime-Cam photometry to compute $d_7$ as follows: (1) The $V-R$ vs. $R-z^{\prime}$ and $R-i^{\prime}$ vs. $i^{\prime}-z^{\prime}$ colors for cluster members were used to derive the locus of galaxy colors at the redshift of the cluster by fitting a 2nd order polynomial to the members in each color-color diagram.  (2) Galaxies in the Suprime-Cam photometry catalog with colors within the 3$\sigma$ scatter about these loci were used to construct the multi-color-selected list of galaxies used for computing local densities.  This multi-color selection served to efficiently subtract contaminating background galaxies.  (3) We excluded galaxies that were fainter than $M_{V_z}=-20.45$.  This magnitude limit corresponds to $M_{V}^*(z=0.83)+1.5$, where $M_{V}^*(z=0.83)=M_{V}^*(z=0)-0.8z$.  This is the same depth in the luminosity function used by previous authors in computing local densities, see \citet{holden2007} and references therein.  We followed \citet{postman2005} in using $M_{V}^*(z=0)=-21.28$, and in implementing a redshift dependent correction to account for passive evolution in the characteristic magnitude of the luminosity function between $z=0.834$ and $z=0$.  The magnitude limit is also very close to $M_{V_z}$ for a red-sequence galaxy at our mass limit.  Local density contours are shown in gray in Figure~\ref{map0152} for $\Sigma =$ 10, 20, 30, 60, and 100~\galmpc.

The multi-color-selected list represents a combination of galaxies with redshifts that are in the cluster redshift interval and contaminants that are not.  A subset of galaxies in the multi-color-selected list have redshifts measured from high-resolution spectroscopy (\zspec).  An analysis of this subset reveals that contaminants with redshifts outside of the interval, \zwindow, account for $\sim 16\%$ of the list.  After assigning a redshift (\zspec\ or \zldp), when available, to objects in the multi-color-selected list, we find that 18\% of galaxies contain at least 1 contaminant within their 7-th nearest neighbor circles.  Furthermore, only $\sim$2\% of galaxies have more than 50\% of the objects within their 7-th nearest neighbor circles identified as contaminants.  We conclude that for the vast majority of galaxies in our study, local density measurements are not significantly impacted by contaminants that exist in the multi-color-selected list, therefore we do not remove the small number of known contaminants since the high-resolution spectroscopic data do not uniformly cover the entire field.

\subsection{Virial Radii} \label{virialradii}
In this work, we study the properties of galaxies inside and outside of the two cores.  The boundaries are given by their virial radii, $r_{200}$, the radius within which the average density is 200 times the critical density of the universe at $z=0.834$.  We use the velocity dispersions of the two cores to determine $r_{200}$ with the relation given in \citet{carlberg1997}:  
\begin{eqnarray}
r_{200}&=&\frac{\sqrt{3} \sigma}{10 H(z)} \nonumber
\end{eqnarray}
Here, $\sigma$ is the velocity dispersion, $H(z)$ is the Hubble constant at the cluster redshift and $r_{200}$ is the virial radius one would measure given three spatial dimensions of information.  The observations however rely on 2-D information.  When projected onto a 2-D plane, the average radius of a ray with length $r$ emanating from the center of a sphere to its surface is equivalent to $R^p = (2/\pi) r$ \citep{limber1960}.  As a result, we define the observed virial radius as:
\begin{eqnarray}
R_{200}^p&=&\frac{2}{\pi} r_{200} \nonumber
\end{eqnarray}
We use the established convention where lowercase letters denote physical 3-D quantities while capitalized ones represent projected values on a 2-D surface.  The superscripted `$p$' in the expression above further implies that the spherical projection factor of ($2/\pi$) has been applied.  Note that this definition for the virial radius is different from most other works.  It is used to minimize contamination in the two cores from non-virialized portions of the halos and field interlopers, which make up an increasing portion of the sample at larger clustercentric radii.

The velocity dispersions of the northern and southern cores are $\sigma=~$768~\kmps\ and $\sigma=~$408~\kmps\ respectively \citep{girardi2005}, which imply projected virial radii of \rtwoN~=~0.76~Mpc and \rtwoS~=~0.40~Mpc.  These radii are represented by the two overlapping inner cyan-colored circles in Figure~\ref{map0152}.  We define $R_{200}^p$ to mark the boundary given by the virial radii of the two cores and therefore when discussing galaxies at clustercentric radii $R<R_{200}^p$ we are referring to those galaxies within both cores.

For calculating distances from the centers of these two cores, we use their Chandra X-ray centroids: ($\alpha,\delta$)~=~(1:52:44.18, -13:57:15.84) for the northern clump and ($\alpha,\delta$)~=~(1:52:39.89, -13:58:27.48) for the southern clump \citep{maughan2003}.  The central position of the cluster as a whole lies between the two cores and is taken to be ($\alpha,\delta$)~=~(1:52:41.669, -13:57:58.32) \citep{girardi2005}.  The clustercentric radius for galaxies, $R$, is defined from this position.  All ($\alpha,\delta$) are in J2000.0 coordinates.

\section{Results} \label{results}
In the following sections we examine the colors of galaxies in different radial and density regimes of \rxj.  The two known AGN detected in X-ray observations \citep{demarco2005} are not included in our analysis because their optical colors are most likely contaminated.

\subsection{Galaxy Colors as a Function of  Clustercentric Radius} \label{colorradius}

\begin{figure}
\epsscale{1.2} 
\begin{center}
\plotone{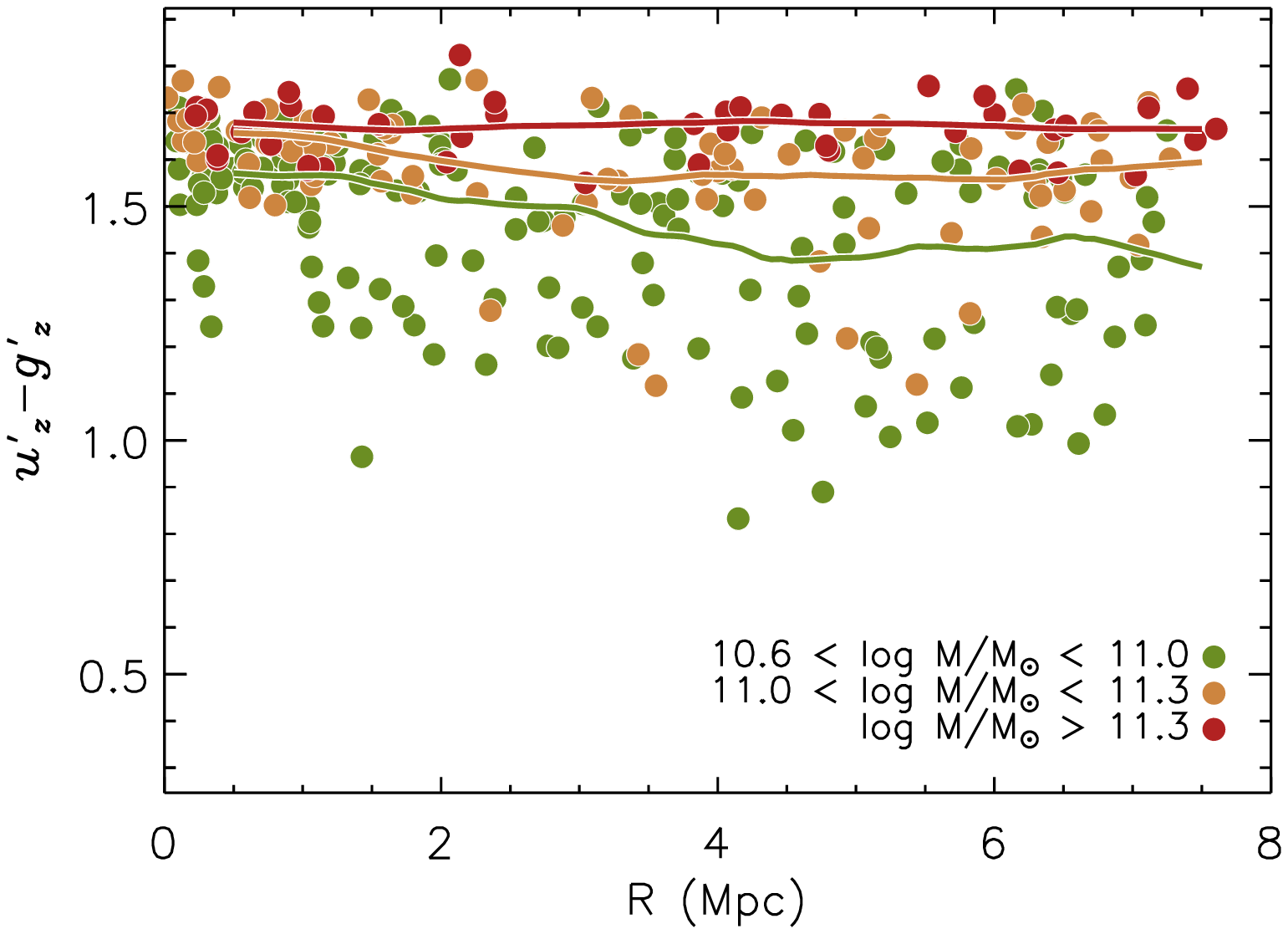}
\caption{Rest-frame \ugz\ color as a function of projected clustercentric radius.  Red points represent galaxies with mass \hirange, orange points \midrange\ and green points \lorange.  Note that most of the data points for the lowest mass galaxies at $R \la 1$~Mpc are not visible but are located on the red-sequence.  The colored lines correspond to a smoothed median color for galaxies in these three mass bins.  The highest mass galaxies are red at all radii while lower mass galaxies span a broader range in colors at $R \ga 1$~Mpc.  Given that galaxies in the outskirts will fall into the cluster core, where almost all galaxies are red, many will have to transition from bluer colors, particularly galaxies in the mass range \lorange.}\label{crr}
\end{center}
\end{figure}

\begin{figure}
\epsscale{1.2} 
\plotone{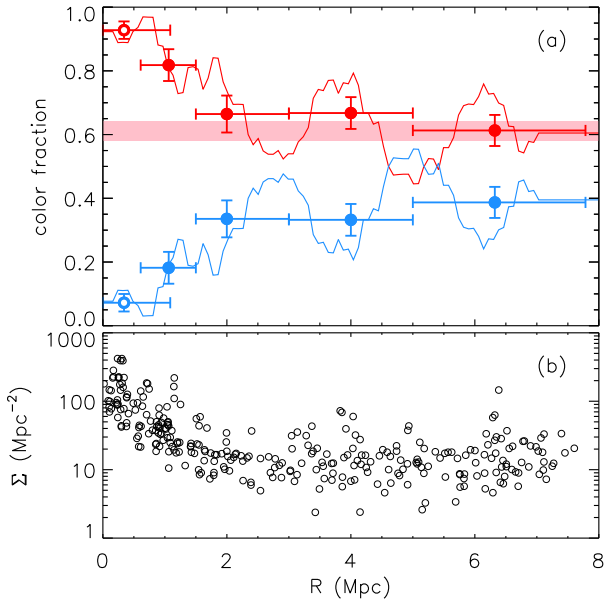}
\caption{Fraction of galaxies with mass \mcutrange\ on the red-sequence ({\em red circles}) or in the blue cloud ({\em blue circles}) as a function of projected clustercentric radius ($a$).  Open circles represent galaxies within the two cores.  Colored lines represent the color fraction at a given radius for the nearest 30 galaxies.  The blue and red galaxy trends are complementary, since all galaxies are defined to be one of the two.  The shaded band indicates a red galaxy fraction of $61 \pm 3\%$ for a $z \sim 0.8$ field sample from \citet{vanderwel2007b} adjusted for the rest-frame \ug\ color used in this work.  The red galaxy fraction is $>90\%$ in the two cores and drops to a level consistent with that of the field at $R \ga 3$~Mpc. This figure reinforces what we see in Figure~\ref{crr}, namely that the two merging cores are dominated by red galaxies, while the outer regions have more blue galaxies.  As galaxies at larger radii fall into the central regions of the cluster, the blue population must be transformed into a red one.  In the bottom panel ($b$), we plot local galaxy density ($\Sigma$) vs. clustercentric radius.  Note that for $R \la 2$~Mpc there is a correlation between $\Sigma$ and $R$.  At larger radii however, the correlation is non-existent.  Consequently, at these larger radii ($R \ga 3$~Mpc) one can isolate the impact of the local environment (i.e. $\Sigma$) on galaxy properties.  For example, the fluctuations in the color fractions in ($a$) at $R > 3$~Mpc may be due to the density peaks at $R \sim 4$~Mpc and $R \sim 6$~Mpc seen in ($b$).} \label{redseqfrac}
\end{figure}

In Figure~\ref{crr} we show \ugz\ color plotted against projected clustercentric radius for galaxies in three mass bins above \mcutrange.  The colored lines represent smoothed median colors for galaxies in these three mass bins.  High mass galaxies lie on the red-sequence at all radii, while lower mass galaxies show bluer colors at larger radii where the local density is lower.  This suggests that as galaxies fall into the high density cluster environment, where most galaxies above our mass limit are red, any evolution that takes place occurs in the lower mass galaxies in our sample.

\begin{figure*}
\epsscale{1.15}
\plottwo{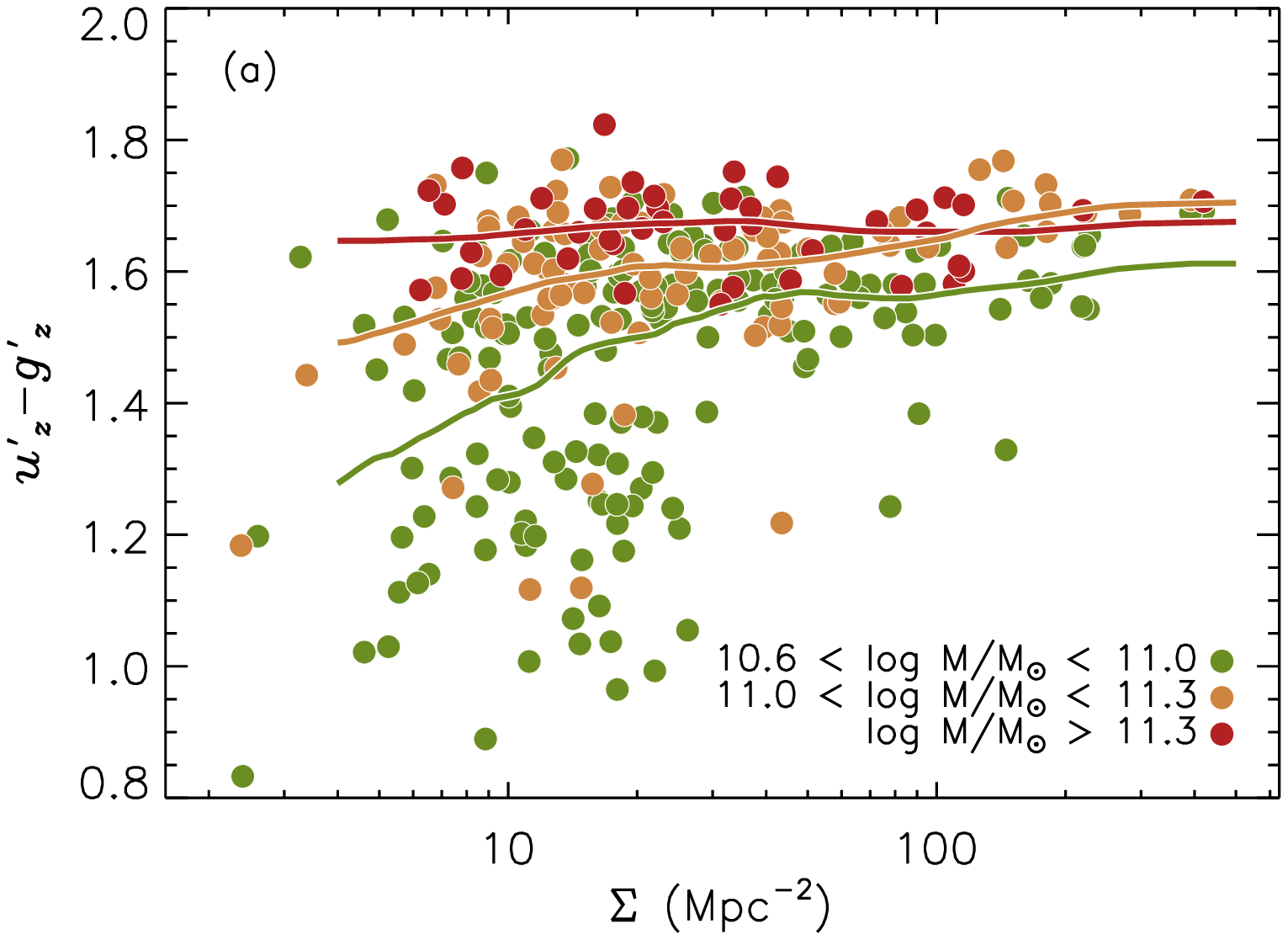}{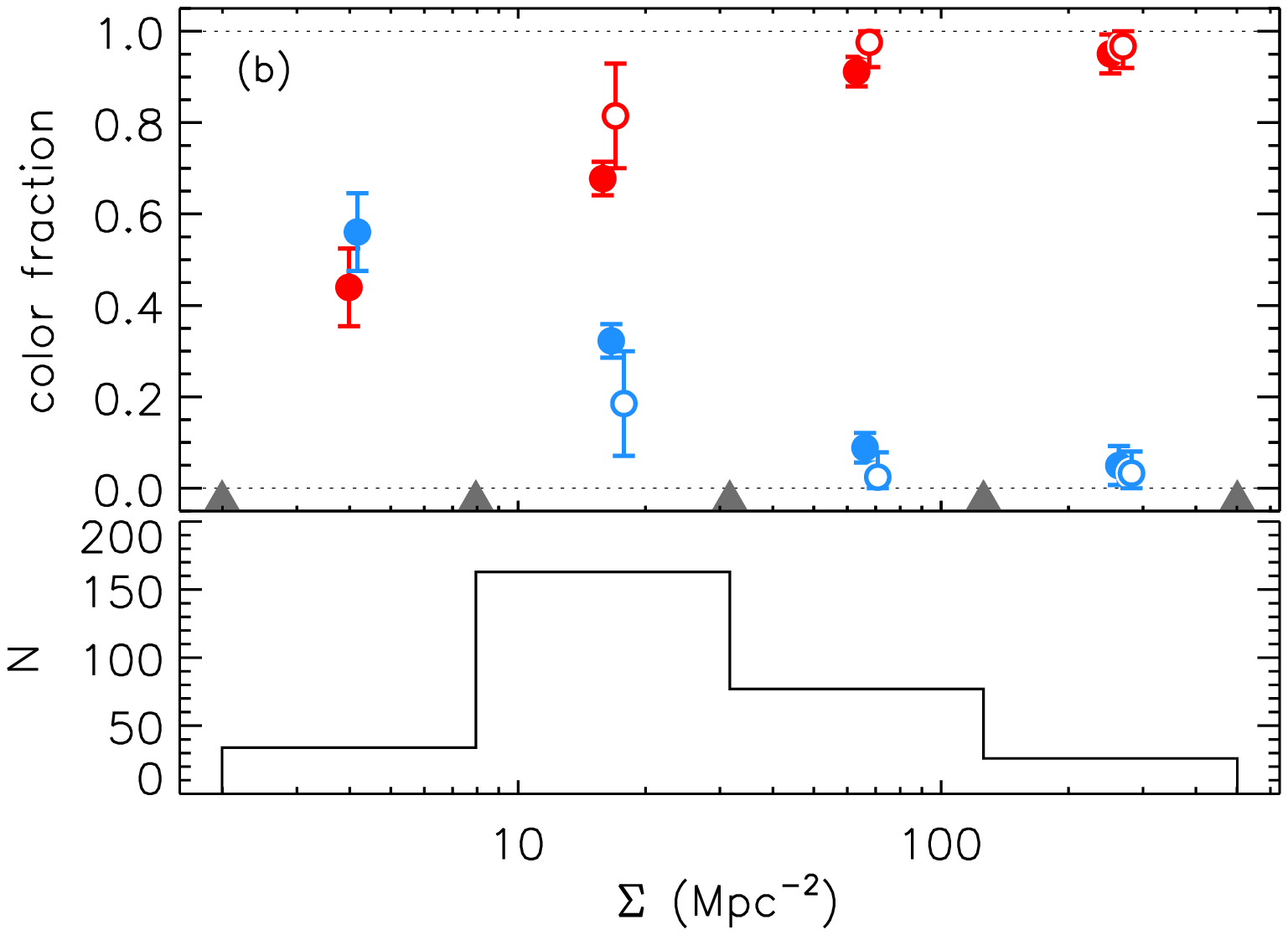}
\caption{Rest-frame \ugz\ color vs. local density for galaxies with mass \mcutrange\ in \rxj\ and its outskirts ($a$).  The color coding is the same as in Fig.~\ref{crr}: red points represent galaxies with mass \hirange, orange points \midrange\ and green points \lorange.  The solid colored lines correspond to a smoothed median color for galaxies in the three mass bins.  The highest mass galaxies are on the red-sequence at all densities while lower mass galaxies have a broader range in color at lower densities (cf. Figure~\ref{crr}).  The fraction of galaxies with red or blue colors as a function of local density is shown in ($b$).  Solid circles are observed color fractions while open circles represent the color fractions after statistically subtracting field interlopers from the sample. We use galaxies in the lowest density bin at $R>3$~Mpc for the field interloper sample.  For this reason, no corrected value is shown for the lowest density bin in this sample.  The number of galaxies in each density bin is indicated at the bottom of the figure.  The increasing fraction of red galaxies with local density shows the expected correlation between local density and clustercentric radius at $R<2$~Mpc.} \label{colordensityall}
\end{figure*}

In Figure~\ref{redseqfrac} we show the fraction of galaxies with red or blue colors as a function of projected clustercentric radius for the mass-limited sample.  The radial bin sizes are indicated by the horizontal error bars and were chosen to provide roughly uniform signal-to-noise in each bin.  The projected radius of each data point represents the median for a particular bin.  Error bars were computed using a binomial distribution.  Note that the apparent overlap between the two smallest radial bins is a consequence of assigning all galaxies within the two virialized cores, which extend to different {\em clustercentric radii}, to a single data point ({\em open circle}).  Galaxies are not being double counted in these two radial bins.  The colored lines represent the color fraction at more highly sampled radial intervals.  They are computed using a sample of the 30 nearest galaxies at a given radius.  The uncertainty in the color fractions for these highly sampled radial intervals is $\sim$9\%.  Note that given our definition of galaxy color, blue galaxies will exhibit a complementary trend to red galaxies and are plotted here for illustrative purposes.  \citet{vanderwel2007b} find a red-sequence fraction for the GOODS field sample at $z \sim 0.8$ of $61 \pm 3\%$ when adjusted to rest-frame \ug\ color (as opposed to $U-B$ in their work).  The shaded red band in Figure~\ref{redseqfrac} represents their result.

Within the two virialized cores, the red galaxy fraction is $f_R = 93 \pm 3\%$.  Moving to larger clustercentric radii, the fraction of red galaxies decreases until $R \sim 2-3$~Mpc.  Beyond $R>3$~Mpc the red fraction is roughly constant at $64 \pm 3\%$.  This relatively high fraction highlights the importance of stellar mass in galaxy evolution at $z \sim 0.8$; above our mass limit, the majority of galaxies that will fuel the future growth in mass of the cluster are already red.  The value for the red galaxy fraction at $R>3$~Mpc is consistent with the field measurement of \citet{vanderwel2007b}, implying that a field population is reached at $\sim 3-4$ times the virial radius of the more massive northern clump, \rtwoN.  \citet{hansen2007} analyzed a sample of clusters at low redshift and also found that the red galaxy fraction reaches a constant value at similar distances.

The red line representing the red fraction at highly sampled radial intervals in Figure~\ref{redseqfrac} reveals localized regions of high or low red galaxy fractions.  This can also be seen by eye in Figure~\ref{crr}.  The peaks in the red galaxy fraction at $R>3$~Mpc occur at $R \sim 4$~Mpc and $R \sim 6$~Mpc.  These are the clustercentric radii of overdense regions identified as groups by \citet{tanaka2006}.  Panel ($b$) of Figure~\ref{redseqfrac}, which shows local galaxy density vs. clustercentric radius, confirms the presence of over-densities at these radii.  In the next section we argue that the elevated red galaxy fraction at these radii is a consequence of the local environment.

\subsection{Galaxy Colors as Function of  Local Density} \label{colordensity}

\begin{figure*}
\epsscale{1.15}
\plottwo{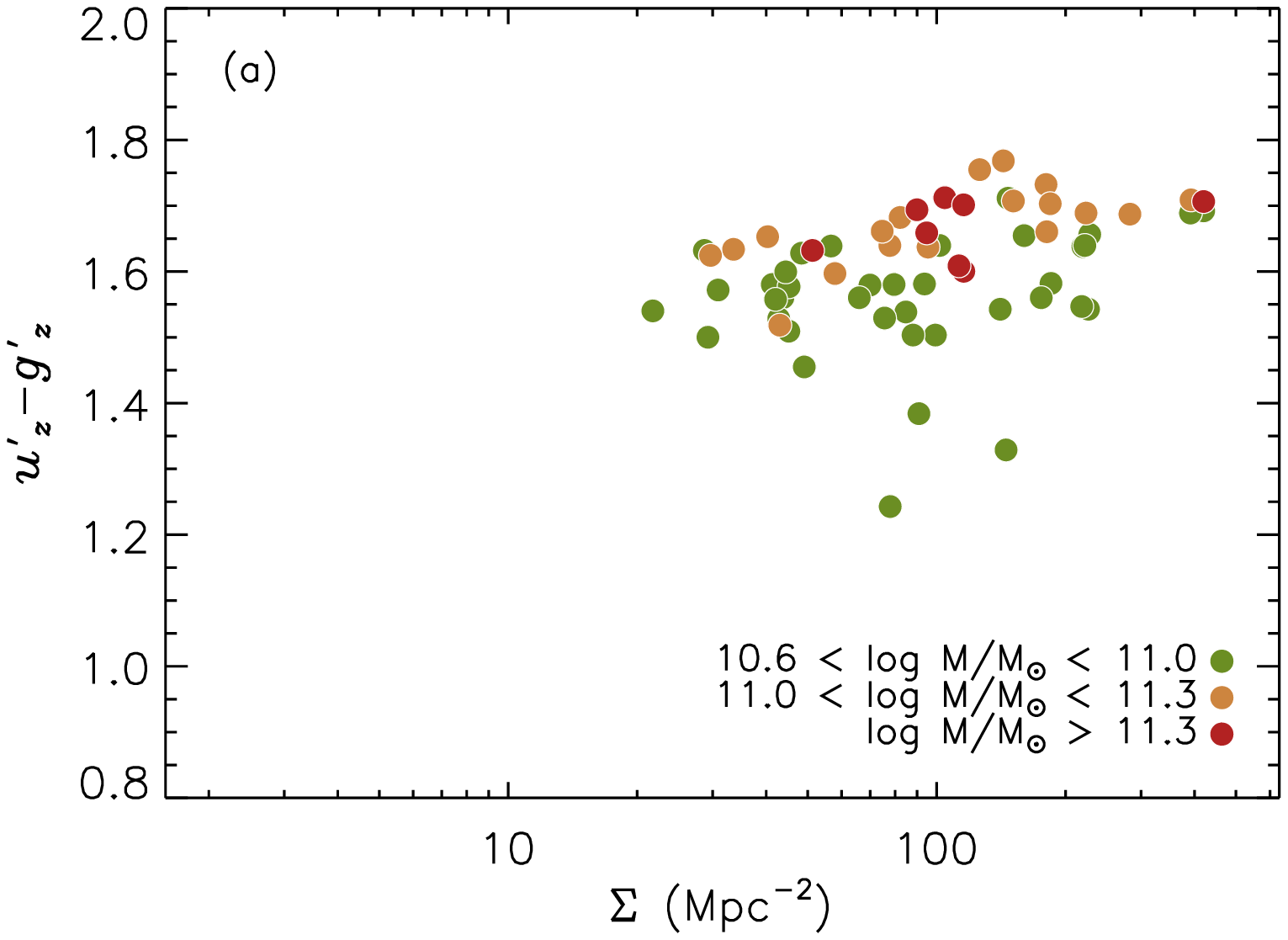}{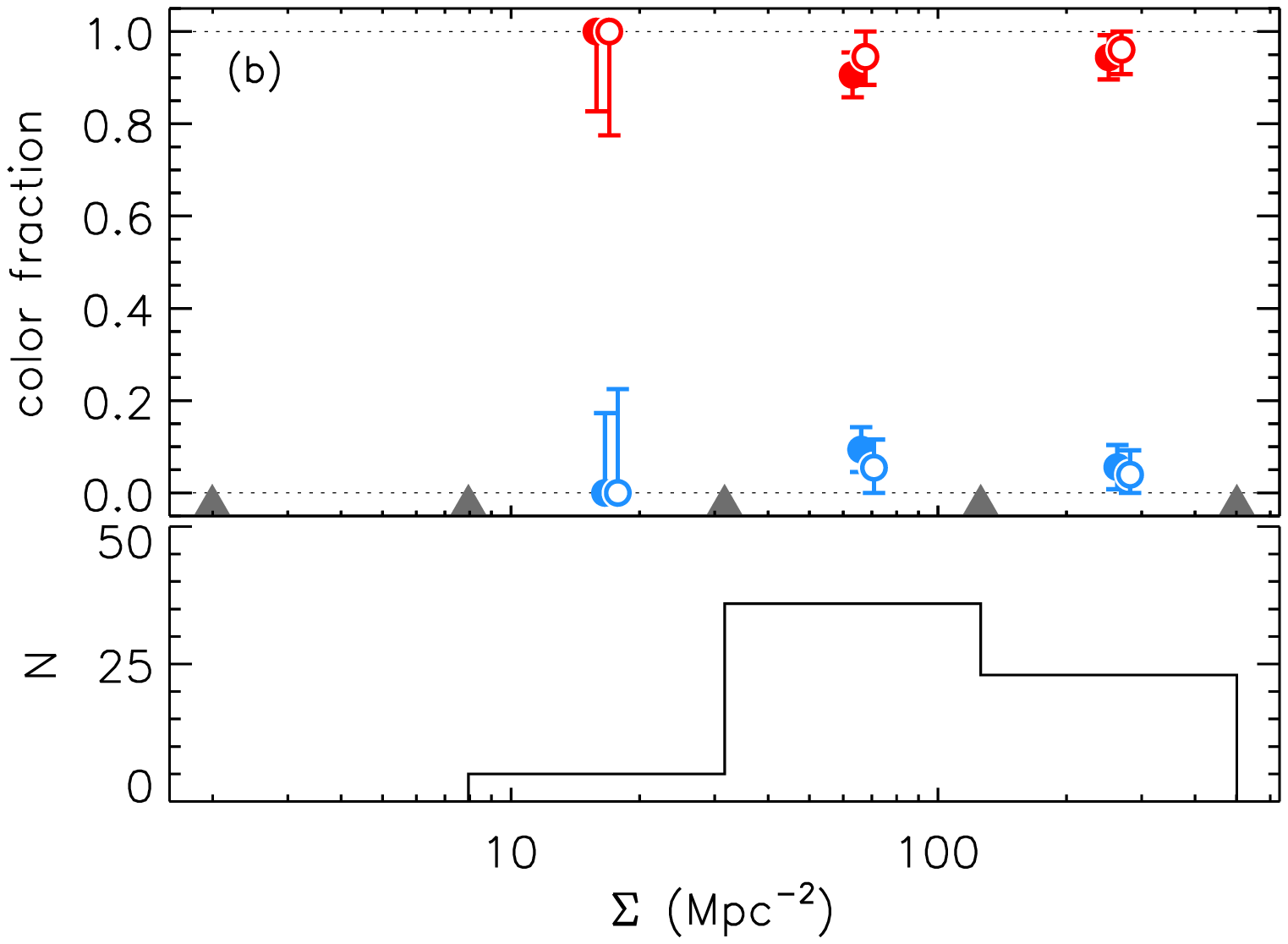}
\caption{Same as Figure~\ref{colordensityall} but for galaxies in the two cores ($R<R_{200}^p$).  Data points for density bins with less than 3 galaxies are not plotted.  A correction for field interlopers was made using the lowest density bin at $R>3$~Mpc (as was also done in Figure~\ref{colordensityall}$b$).  Very few galaxies above our mass limit have blue colors inside of the two cores (see ($a$)) and therefore the red fraction is near $\sim 100\%$ at all densities, as seen in ($b$).  We rule out a correlation between color and density in ($a$) after computing a Pearson correlation coefficient of $\sim 0.35$.} \label{colorfracdensity1}
\end{figure*}


In this section we analyze the red and blue galaxy fractions as a function of local density.  In doing so, we correct the color fractions for contamination from field interlopers, i.e. those with redshifts within our chosen window (\zwindow), but which lie in the line-of-sight far from the cluster.  In the previous section, we found that the fraction of red galaxies reached the value found in field surveys at $R \ga 3$~Mpc.  We therefore use those galaxies with $R>3$~Mpc and densities of $2~{\rm Mpc^{-2}} < \Sigma < 8~{\rm Mpc^{-2}}$ to form the field interloper sample (see Appendix for details).  The low density criterion is necessary in order to assess the level of contamination in intermediate and high density enhancements at $R>3$~Mpc.  The ``field'' value for the red galaxy fraction, $45 \pm  10$\%, is subsequently used in the discussion, figures and table below.

\begin{figure*}
\epsscale{1.15}
\plottwo{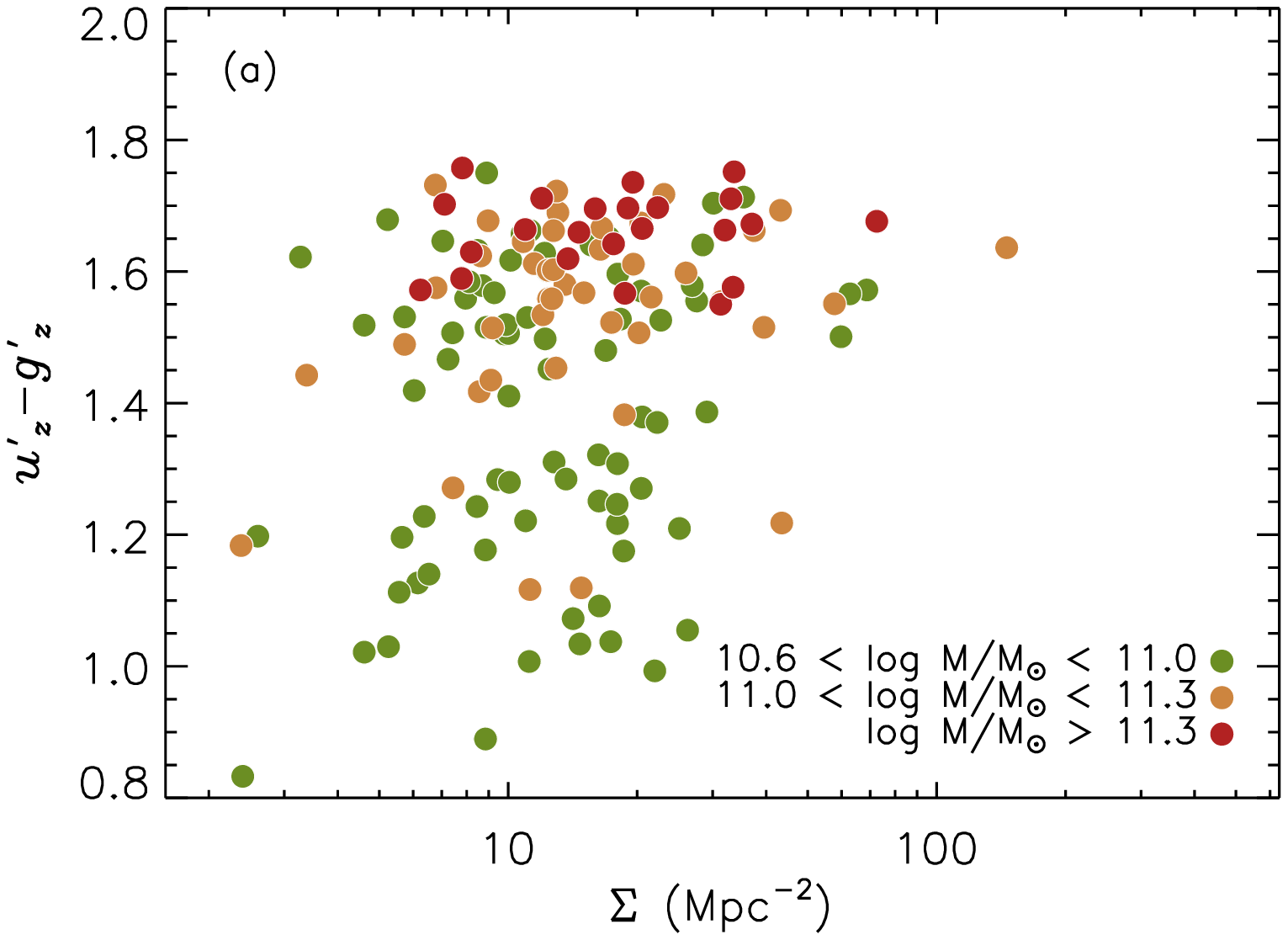}{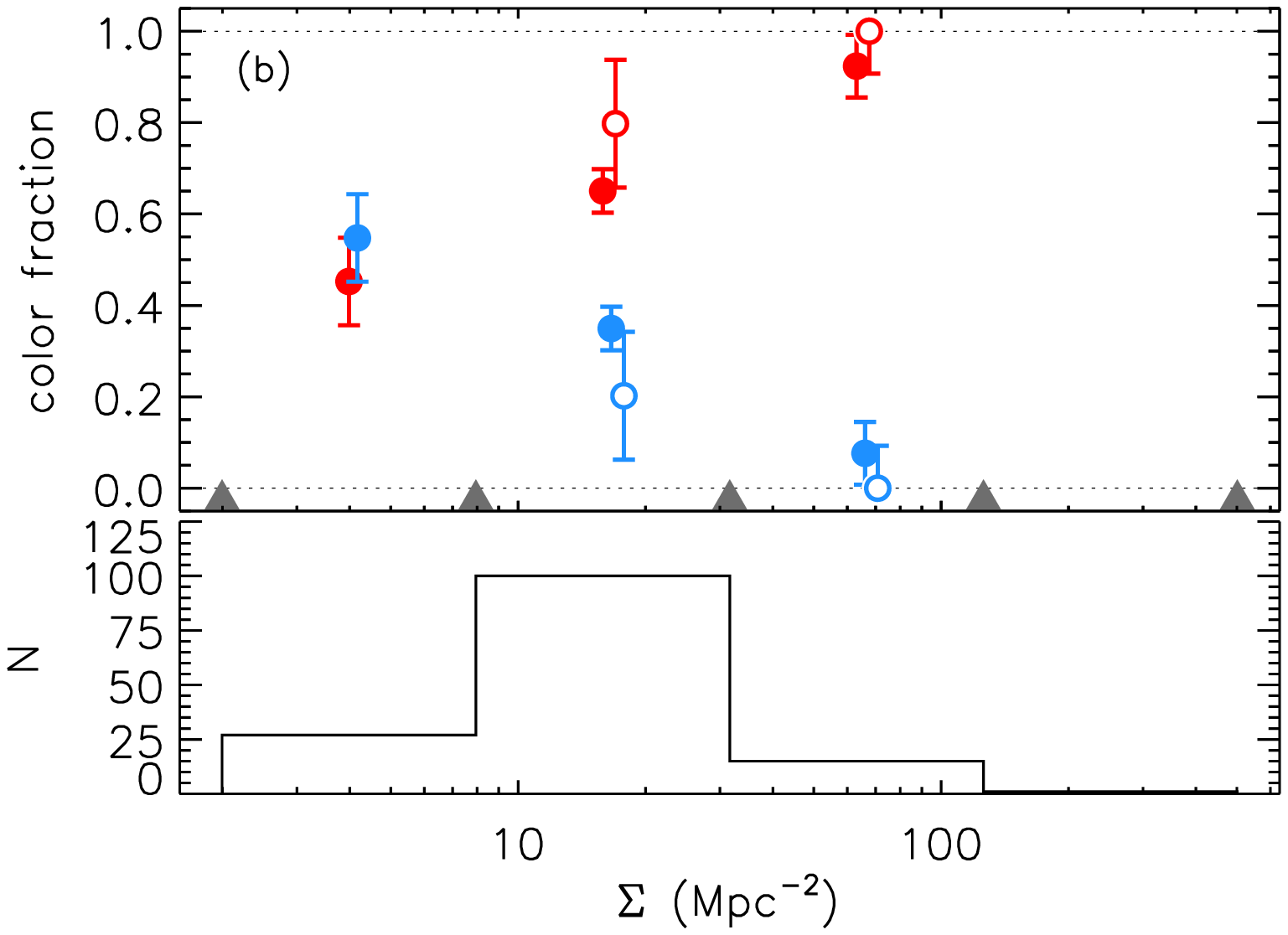}
\caption{Same as Figure~\ref{colordensityall} but for galaxies at large clustercentric radii ($R>3$~Mpc).  The red fraction in the lowest density bin ($45 \pm 10\%$) is taken, as before, to be the field for correcting color fractions at higher densities (see \S~\ref{colordensity} and Appendix).  Galaxies at $R>3$~Mpc are outside the radial regime where clustercentric radius and density are correlated; any dependence of the color fractions on density must be due to local environmental processes instead of processes associated with the cores of clusters.  A large proportion of lower mass galaxies in this sample have blue colors while higher mass galaxies are overwhelmingly red, as seen in ($a$).  The increasing fraction of red galaxies with local density, as shown in ($b$), suggests that the local environment is an important factor in evolving galaxies in the outskirts.  Galaxies at the highest densities at these radii mostly reside in the F4 and F6 groups (see Figure~\ref{map0152}).  In those environments, the red fraction is the same as that of the two cores.  Strikingly, at intermediate densities ($8~{\rm Mpc^{-2}} < \Sigma < 32~{\rm Mpc^{-2}}$), the red fraction is also enhanced relative to the lowest density bin.  These galaxies are not associated with any particular density enhancement, but rather trace the filamentary structure surrounding the cluster.} \label{colorfracdensity45}
\end{figure*}

In Figure~\ref{colordensityall}$a$, we show \ugz\ color vs. local density for galaxies with mass \mcutrange.  As expected from the red points in Figure~\ref{crr}, the highest mass galaxies lie on the red-sequence at all densities, while we see more clearly here the result that was hinted at in Figure~\ref{crr}, namely that galaxies of lower mass exhibit a broader range in colors at lower densities.  Also shown in Figure~\ref{colordensityall}$b$ are the red and blue galaxy fractions vs. local density.  Both the observed color fractions ({\em solid circles}) and the field interloper-corrected fractions ({\em open circles}) are shown.  The red galaxy fractions are given in Table~\ref{table_colordensity}.  A clear trend of galaxy colors exists with local density as expected from the correlation between clustercentric radius and local density at $R \la 2$~Mpc where a large number of galaxies in our sample lie.  At low density there is a roughly equal mix of red and blue galaxies, while at the highest densities, red galaxies are dominant.  This trend of an increasing fraction of red galaxies with density is consistent with what has been seen in previous studies of color trends with density at low redshift \citep{baldry2006}.  

The color-density relation in Figure~\ref{colordensityall}$b$ is an analog of the mass-limited morphology-density relation \citep{holden2007,vanderwel2007b}; which is related to the \citet{dressler1980} luminosity-based MDR.  As is generally recognized, galaxy colors and morphology track one another (see e.g. \S~\ref{colorfractions}).  \citet{holden2007} computed the mass-limited MDR for clusters spanning redshifts of $0<z<1$ for the same stellar mass-limit as in this work.  The mass-limited MDRs in that work showed weak correlations between early-type fraction and density.  Our survey reaches to larger clustercentric radii and so we are able to probe galaxy properties to much lower local densities.  Interestingly, we find a stronger dependence of red galaxy fraction on local density when including densities that are lower than what could be investigated in \citet{holden2007}.  This is consistent with the mass-limited MDR of \citet{vanderwel2007b}, which shows a large range in early-type fraction with density when combining their low density field sample with the high density cluster sample of \citet{holden2007}.  The large range in red galaxy fractions seen in our uniformly selected sample therefore confirms the morphological trends seen in the work of \citet{vanderwel2007b}.

\begin{deluxetable*}{lcccc} 
\tablewidth{0pc}
\tablecolumns{5}
\tablecaption{Red Galaxy Fractions\tablenotemark{a}\tablenotemark{b}\tablenotemark{c} \label{table_colordensity}}
\tablehead{
\colhead{$R$} & \colhead{$2~{\rm Mpc^{-2}} < \Sigma < 8~{\rm Mpc^{-2}}$} & \colhead{$8~{\rm Mpc^{-2}} < \Sigma < 32~{\rm Mpc^{-2}}$} & \colhead{$32~{\rm Mpc^{-2}} < \Sigma < 126~{\rm Mpc^{-2}}$} & \colhead{$126~{\rm Mpc^{-2}} < \Sigma < 501~{\rm Mpc^{-2}}$}
}
\startdata
All & $0.44_{-0.09}^{+0.09}$ (34) & $0.68_{-0.04}^{+0.04}$ (163) & $0.91_{-0.03}^{+0.03}$ (77) & $0.95_{-0.04}^{+0.04}$ (26)  \\
\nodata (corr) & \nodata & $0.81_{-0.11}^{+0.11}$\phm{ (000)} & $0.98_{-0.05}^{+0.02}$\phm{ (00)} & $0.97_{-0.05}^{+0.03}$\phm{ (00)}  \\
$R<R_{200}^p$ & \nodata & $1.00_{-0.17}^{+0.00}$ (5)\phm{00} & $0.91_{-0.05}^{+0.05}$ (36) & $0.94_{-0.05}^{+0.05}$ (23)  \\
\nodata (corr) & \nodata & $1.00_{-0.22}^{+0.00}$\phm{ (000)} & $0.95_{-0.06}^{+0.05}$\phm{ (00)} & $0.96_{-0.05}^{+0.04}$\phm{ (00)}  \\
$R>3$~Mpc & $0.45_{-0.10}^{+0.10}$ (27) & $0.65_{-0.05}^{+0.05}$ (100) & $0.92_{-0.07}^{+0.07}$ (15) & \nodata  \\
\nodata (corr) & \nodata & $0.80_{-0.14}^{+0.14}$\phm{ (000)} & $1.00_{-0.09}^{+0.00}$\phm{ (00)} & \nodata  \\
\enddata
\tablenotetext{a}{The first row of each radial regime gives the observed red galaxy fraction while the second row gives the value that is corrected for contamination from field interlopers.  Note that no correction is applied to the lowest density bin.  The red fraction for the field is assumed to be $45 \pm 10\%$.}
\tablenotetext{b}{The raw number of galaxies in each density bin is given in parenthesis.}
\tablenotetext{c}{Red galaxy fractions are computed only for density bins with 3 or more galaxies.}
\end{deluxetable*}

In order to gauge the role of local environment on galaxy evolution, we separate galaxies into two groups, as shown in Figures~\ref{colorfracdensity1} and \ref{colorfracdensity45}.  The first sample includes galaxies in the two cores ($R<R_{200}^p$) while the second is those at larger clustercentric radii ($R>3$~Mpc).  The sample at $R>3$~Mpc is far enough away from the region where density and clustercentric radius are correlated ($R \la 2$~Mpc, see Figure~\ref{redseqfrac}$b$) that local environmental effects (e.g. galaxy-galaxy interactions, etc.) can potentially be untangled from more global processes associated with clusters (e.g. ram-pressure, harassment, etc.).  This is a particularly important benefit and feature of studying galaxy evolution at large radii, yet still in the cluster environment.  Galaxies in the outskirts will, with high likelihood, be those that ultimately populate the core and will therefore be progenitors of typical cluster red-sequence galaxies.  The red galaxy fractions for both radial regimes are summarized in Table~\ref{table_colordensity}.

In the two cores, the red galaxy fraction is near unity for all densities (Figure~\ref{colorfracdensity1}$b$).  This is true regardless of the stellar mass of the galaxy (Figure~\ref{colorfracdensity1}$a$).  At larger clustercentric radii ($R>3$~Mpc), galaxies obviously populate lower density environments (Figure~\ref{colorfracdensity45}).  It is a fascinating result that high mass galaxies in this radial regime are predominantly red regardless of density, while a large fraction of lower mass (but still massive) galaxies are found to have blue colors in the two lowest density bins (Figure~\ref{colorfracdensity45}$a$).  When considering all galaxies above our mass limit at $R>3$~Mpc, we find an increasing fraction of red galaxies towards higher densities.  For galaxies that occupy the highest density regions ($32~{\rm Mpc^{-2}} < \Sigma < 126~{\rm Mpc^{-2}}$), the red galaxy fraction is similar to the value found in the two cores for the same density range.  This suggests that galaxies in high density environments in the outskirts of clusters have already evolved by $z \sim 0.8$ to be similar, and to occur in similar proportions, to galaxies that occupy the two cores.

The presence of an increasing red galaxy fraction trend with local density implies that certain environments nurture or drive the transition from blue to red at a level above what is found in the field.  The red galaxy fraction in the density range $8~{\rm Mpc^{-2}} < \Sigma < 32~{\rm Mpc^{-2}}$ at $R>3$~Mpc is between the value found in adjacent density bins: in the higher density bin, most galaxies are already evolved to the canonical core fraction while in the lower density bin, we have reached the field population level where the red fraction is low (at $45 \pm 10\%$ for our sample).  The density range $8~{\rm Mpc^{-2}} < \Sigma < 32~{\rm Mpc^{-2}}$ therefore represents a regime where one expects the local environment to influence evolution above a level that is typical for the field and where star formation begins to shut off in galaxies, causing them to move towards redder colors.  Given the large number of galaxies at these densities, a potentially large sample exists of galaxies that are beginning to transition from blue to red in these particular environments.  This sample will be a major resource for assessing the physical processes that drive galaxies from field-like galaxy properties to cluster-like galaxy properties.  Additional observations to complement galaxy colors, such as those that allow morphological identification, estimates of star formation rates, structural measurements and assessments of merger events would be useful in characterizing such galaxies.

\section{Discussion} \label{discussion}
The simulations of \citet{wechsler2002} suggest that massive clusters like \rxj\ will grow by more than a factor of two in mass between $z \sim 0.8$ and today.  Such massive clusters formed out of the highest density perturbations from the primordial fluctuations in the early universe.  As a consequence, these high density perturbations were surrounded by intermediate density perturbations that have collapsed to form structures on the group mass scale that are subsequently accreted into the cluster.  Therefore, much of the mass buildup in massive clusters like \rxj\ occurs through the accretion of nearby infalling groups.  The N-body simulations of \citet{berrier2008} however, suggest that clusters grow primarily through the accretion of individual galaxies from the field.  However, they also note that for more massive clusters, such as \rxj, a larger proportion of galaxies are accreted as part of group sized halos.

\subsection{Evolution in Infalling Groups}

Most of the galaxies above our mass limit at $R>3$~Mpc and at high densities ($32~{\rm Mpc^{-2}} < \Sigma < 126~{\rm Mpc^{-2}}$) lie in two groups identified by \citet{tanaka2006}.  As in the two cluster cores, the fraction of red galaxies in these groups is $\sim 90-100\%$.  One group is located at a projected separation of $8\arcmin$ ($\sim 4$~Mpc) south of the cluster center and the other $13\arcmin$ ($\sim 6$~Mpc) southwest of the cluster center (see Figure~\ref{map0152}).  \citet{tanaka2006} refer to these systems as F4 and F6 respectively and measure velocity dispersions of $413 \pm 241$~\kmps\ and $457 \pm 46$~\kmps.  Their offsets in redshift from the cluster ($z_{\rm F4}-z_0 \sim 0.010$ and $z_{\rm F6}-z_0 \sim 0.001$) imply relative velocities ($v=c(z-z_0)/(1+z_0)$) of $\sim 160$~\kmps\ and $\sim 1600$~\kmps, which means that both are encompassed by the $\sim 2\sigma$ velocity dispersion of the northern core.  If their redshift offsets instead are due to expansion with the Hubble flow, the line of sight distance ($d=c\Delta z/H(z)$) from the cluster would be $\sim 3$~Mpc and $\sim 30$~Mpc respectively.  The F4 group is therefore very likely to be falling into the cluster and given that it has a projected distance from the cluster center of $\sim 4$~Mpc, it wouldn't be unreasonable for the F6 group, with a projected distance of $\sim 6$~Mpc, to also be falling into the cluster, though this outcome is less certain.

Evidence that the source of the constituent galaxies of massive clusters (like \rxj) are groups in the outskirts has also been seen in previous work.  \citet{dressler1997} found that assembling clusters at intermediate redshift had a high fraction of elliptical galaxies and was similar to the fraction found in virialized clusters, suggesting that many ellipticals were in place prior to virialization of massive clusters and probably formed earlier, in the group phase.  At $z=0.37$, \citet{gonzalez2005b} found that at least four gravitationally bound groups were in the process of collapsing to form a larger cluster.  A higher proportion of galaxies in these groups were found to be passive when compared to the field, similar to our findings.  Others have also found groups or overdense regions in the outskirts of clusters to support galaxies with properties that are similar to those found among galaxies in cluster cores.  For instance, \citet{moran2007b} found passive spirals in the outskirts of a $z \sim 0.5$ cluster to be preferentially located in overdense regions, perhaps in transition to becoming S0s.  Meanwhile, \citet{treu2003} found the morphological mix of overdense regions in the outskirts of a $z \sim 0.4$ cluster to be very similar to the central regions.

Regardless of their proximity to the two main cores of \rxj, the groups in the redshift interval \zwindow, contain an enhanced fraction of red galaxies.  The velocity dispersions of the F4 and F6 groups imply virial radii, $r_{200}$ (i.e. unprojected), of $\sim 0.6$~Mpc and $\sim 0.7$~Mpc respectively.  The median local density of galaxies in these groups is $\Sigma \sim 40$~\galmpc.  Recall that these densities are computed for galaxies above a luminosity-limit ($M_{V_z} < -20.45$).  Converting this surface density into a true space density with the virial radii computed above, we find $n \sim 40-50$~Mpc$^{-3}$.  Given that the relative velocities of the galaxies in these groups are much closer to the galaxy interval velocities, and that the galaxies are in regions of high density, the frequency of galaxy-galaxy interactions and merging is enhanced.  The merger of disk galaxies can lead to the formation of ellipticals \citep{toomre1972, schweizer1982, cox2006b} and processes associated with merger events, such as starbursts \citep{mihos1996, sanders1996} and AGN feedback \citep{springel2005} among others \citep[see][and references therein]{faber2007}, can quench star formation and move galaxies towards the red-sequence.  Thus, mechanisms in these group environments exist to develop and enhance a high red galaxy fraction.

\subsection{Evolution in Intermediate Density Environments}
In intermediate density enhancements ($8~{\rm Mpc^{-2}} < \Sigma < 32~{\rm Mpc^{-2}}$: somewhat lower in density compared to the groups F4 and F6 discussed above) at large clustercentric radii, $R>3$~Mpc, the red galaxy fraction is also elevated relative to the value in the lowest density bin.  These intermediate densities therefore represent a regime in which galaxies begin to transition towards the red-sequence in larger proportions than what is found at the lowest densities, where the local environment is not effective in influencing galaxy evolution.  Galaxies at these intermediate densities are not all obviously identified with large groups in the outskirts such as the F4 and F6 groups, though many are found near them.  Instead, they appear to trace the filamentary structure surrounding the cluster, and so may indicate the sites of the earliest stages of transformation to red, quiescent early-type galaxies.

The local environment of a galaxy can be thought of as a tracer for the underlying mass of the dark matter halo in which it resides.  \citet{conroy2008b} explored abundance matching and found that a galaxy's {\em halo mass} determines the amount of {\em stellar mass} the galaxy will have at $z \sim 1$ as well as its star formation rate.  Their work focuses primarily on the central galaxies in dark matter halos.  The satellite galaxies that occupy subhalos, which are part of more massive group and cluster sized halos, are not specifically addressed in their work.  The ubiquitous red colors of the highest mass galaxies in our sample could reflect the lack of star formation in high mass galaxies found at $z \sim 1$ in this model.  Galaxies of lower mass in the model are expected to have higher star formation rates at this epoch.  The low mass galaxies in our sample that reside outside of the two cluster cores and infalling groups support this view given their blue colors.  However, it is plausible, even likely, that these lower mass galaxies will eventually fall into larger mass halos, where a number of mechanisms are capable of changing their structure and halting star formation, some of which were discussed above.

\section{Summary \& Conclusions} \label{summary}
We studied the environmental dependence of galaxy colors in the $z=0.834$ galaxy cluster \rxj\ in order to investigate where galaxies transition from blue, late-type systems into red, early-type ones that dominate the central regions.  The cluster is a massive X-ray luminous system that is comprised of two cores.  The northern and southern cores have velocity dispersions $\sigma=~$768~\kmps\ and $\sigma=~$408~\kmps, respectively, and overlapping projected virial radii of \rtwoN~=~\rtwoNval\ and \rtwoS~=~\rtwoSval\ and are separated in velocity along the line of sight by $\Delta v \sim 1600$~\kmps.  

With a low-dispersion prism (LDP) on IMACS, the multi-object spectrograph on the 6.5~m Baade (Magellan I) telescope, we have built one of the largest samples of {\em spectroscopically} confirmed members for a cluster at intermediate redshift.  This was accomplished by targeting galaxies with \ifil~$<23.75$~mag within a $D=27$\arcmin\ ($D \approx 12$~Mpc at $z=0.834$) FOV encompassing the two cores.  Prior to this work, $\sim$250 galaxies were known to be members of \rxj\ and its outskirts with redshifts in the range $0.80<z<0.87$.  We have added \nldponly\ new members.  The redshifts derived from the LDP show a scatter of $\sigma = 0.018$, or 1\% in $(1+z)$, about those derived from higher resolution spectra (\S~\ref{redshifts}), allowing us to localize galaxies within the superstructure of the cluster.  

Stellar masses for galaxies were computed using the \citet{bell2003} relation between rest-frame $B-V$ color and $M/L_{B}$ (\S~\ref{masses}).  We used a mass-limited sample of \nmcut\ galaxies with mass \mcutrange\ to conduct our study.  This mass threshold corresponds to the \ifil-band magnitude along the red-sequence at which the spectroscopic completeness is $\sim 50\%$ (\ifil~$\sim 23$).  Because the spectroscopic completeness varies for galaxies of different mass, we applied completeness corrections when necessary (\S~\ref{completeness}).  We assigned galaxies to be blue or red based on their deviation from the red-sequence, with blue galaxies having \ugz\ colors $\sim 0.12$~mag or more below the red-sequence.  

This initial sample provides an opportunity to analyze cluster galaxy properties at intermediate redshift with substantial statistical significance, and is only a glimpse of what will be possible when our survey is complete and we are likely to have over $\sim 2000$~members.

Our conclusions are the following:
\begin{enumerate}
\item The most massive galaxies (\hirange) lie on the red-sequence at all clustercentric radii and all local densities, while less massive galaxies have a broad range in colors farther from the two cluster cores and in lower density environments.  In this massive cluster at $z \sim 0.8$ most of the evolution due to environmental effects is occurring in the lower mass galaxies.
\item The fraction of galaxies that are on the red-sequence is at a maximum in the two cores of the cluster at $93 \pm 3\%$.  This red galaxy fraction declines at larger clustercentric radii until $R \sim 3$~Mpc, beyond which it is roughly constant at $64 \pm 3\%$.  Our lowest density sample at $R > 3$~Mpc ($2~{\rm Mpc^{-2}} < \Sigma < 8~{\rm Mpc^{-2}}$) has a red fraction of $45 \pm 10\%$.  These values are broadly consistent with the red galaxy fraction for the field at $z \sim 0.8$.
\item At $R>3$~Mpc, where the correlation between clustercentric radius and density is nonexistent, we find that the red galaxy fraction increases with local density.  This implies that the local environment is a factor in galaxy evolution in the outskirts of the cluster.
\item The high density environments ($32~{\rm Mpc^{-2}} < \Sigma < 126~{\rm Mpc^{-2}}$) in the outskirts ($R>3$~Mpc) have a red galaxy fraction that is similar to what is found in the two cores.  These galaxies are primarily located in two previously identified groups.
\item A large number of galaxies at $R>3$~Mpc are found in our intermediate density bin ($8~{\rm Mpc^{-2}} < \Sigma < 32~{\rm Mpc^{-2}}$), where the red galaxy fraction is elevated relative to the value found for the lowest density bin ($2~{\rm Mpc^{-2}} < \Sigma < 8~{\rm Mpc^{-2}}$).  These intermediate densities therefore represent a regime where galaxies above our mass-limit first begin to undergo transformations from star forming to quiescent as a result of the local environment.  Most of these galaxies are not associated with either of the large groups in the outskirts, but instead trace the surrounding filamentary structure.
\end{enumerate}

Exploring the outer parts of clusters provides opportunities to identify the regions where galaxies of different masses are growing in mass and/or transforming from blue, star-forming, late-type galaxies to red, quiescent, early-type galaxies.  The uniformity of the colors of the highest mass galaxies at $z \sim 0.8$ suggests that the vast majority of their star formation has taken place already, even at this epoch, while the diverse colors of the somewhat lower mass sample indicate that the transformation process is still underway in the intermediate density regions in the outer parts of \rxj. Studies of galaxies in such regions of transformation promise to provide great insight into the physical processes that drive galaxy evolution and build up the red-sequence population.

\subsection{Future Work}
We are in the process of expanding the present sample of galaxies in \rxj\ with the LDP and we are also targeting another cluster at $z \sim 0.8$, \msfull.  As we move forward, the LDP reductions and SED fitting are being refined to provide better constraints on redshifts and other galaxy properties.  We are also in the process of obtaining $K_s$-band imaging, which will allow us to improve our stellar mass estimates.  When completed, this dataset will yield a wealth of information concerning galaxies in overdense environments in and around clusters at intermediate redshift.  Between the two targeted fields, we expect to find 2000-3000 galaxies in the clusters and their outskirts, allowing us to study the environmental dependence of mass functions with $M^{\star}$ and $\alpha$ to significantly higher precision than at present, possibly as high as $\pm 0.1$~dex and $\pm 0.15$ respectively.  We will also explore stellar population parameters from SED fitting, an analysis of blue galaxies down to low masses, and many other exciting aspects of cluster galaxies at an epoch some 7-8~Gyr ago when clusters were rapidly building up.

\acknowledgments
This research was supported by NASA grant NAG5-7697 and Spitzer grant JPL 1277397.  We wish to acknowledge those who have contributed to the construction and deployment of IMACS as well as Scott Burles for developing the low-dispersion prism, and the PRIMUS collaboration for allowing us to investigate clusters with their hardware.  The combination of these two components resulted in the impressive dataset presented in this work.  The authors wish to recognize and acknowledge the very significant cultural role and reverence that the summit of Mauna Kea has always had within the indigenous Hawaiian community.  We are most fortunate to have the opportunity to conduct observations from this mountain.  This research used the facilities of the Canadian Astronomy Data Centre operated by the National Research Council of Canada with the support of the Canadian Space Agency.

\begin{appendix}

\section{Field Interlopers}

The width of the cluster redshift window, \zwindow, was chosen to accommodate redshift uncertainties for galaxies determined to be members from IMACS spectroscopy.  This window is large enough to allow field interlopers to fall into our sample.  These galaxies have redshifts within the cluster redshift window but are far from the cluster superstructure.  Their deviation in redshift from the cluster's reflects expansion with the Hubble flow rather than an induced peculiar velocity from the nearby massive structure.

To correct the red galaxy fractions for contamination from field interlopers we employ a statistical background subtraction.  For a given surface area of size $A$, the number of galaxies with density $\Sigma$ falling into this patch of sky is given by $\Sigma A$ and the number of field interlopers is given by $\Sigma_{\rm field} A$.  The contamination fraction at a particular density $\Sigma$ is therefore $f_c = \Sigma_{\rm field}/\Sigma$.  We use the median density in a particular density bin to represent $\Sigma$ when computing $f_c$.  The corrected red galaxy fraction, $f_R^{\rm cor}$, and its uncertainty, $\sigma_{f_R^{\rm cor}}$, can then be written in terms of the observed red fraction at a particular density, $f_R$, the red fraction for field interlopers, $f_R^{\rm field}$, and their respective uncertainties as follows:

\begin{eqnarray}
f_R^{\rm cor} &=& \frac{f_R - f_cf_R^{\rm field}}{1-f_c}  \nonumber \\
\sigma_{f_R^{\rm cor}} &=& \frac{\sqrt{ \sigma_{f_R}^2 + f_c^2 \sigma_{f_R^{\rm field}}^2  }}{1-f_c}  \nonumber
\end{eqnarray}

We define a field interloper sample using those galaxies with \mcutrange\ at clustercentric radii $R>3$~Mpc and densities of $2~{\rm Mpc^{-2}} < \Sigma < 8~{\rm Mpc^{-2}}$.  This sample represents galaxies in the lowest density bin in Figure~\ref{colorfracdensity45}$b$ where the red fraction is $f_R^{\rm field} = 45\%$ and the uncertainty is $\sigma_{f_R^{\rm field}}=10\%$.  The median density for this sample is $\Sigma_{\rm field}=6$~\galmpc.  At $R>3$~Mpc, the observed red fraction in Figure~\ref{redseqfrac} is roughly constant, implying that a field population has been reached at these clustercentric radii.  We further select galaxies in the lowest density bin to represent field interlopers because we are interested in determining the level of contamination in intermediate and high density clumps at $R>3$~Mpc.

In some cases, the corrected value for the red fraction exceeds unity, especially when the {\em observed} red fraction is already close to it.  This effectively results from over-subtraction of blue galaxies when the observed number of blue galaxies is small.  Under these circumstances, we assign the corrected red fraction to be $f_R^{\rm cor}=100\%$.

\end{appendix}


\end{document}